\documentclass[11pt]{article}
\pdfoutput=1 
\usepackage[sc]{mathpazo}

\usepackage[margin=1in]{geometry}

\usepackage{latexsym,amsfonts}
\usepackage{amsmath}
\usepackage{amssymb,amsthm}
\usepackage{array}

\usepackage{graphicx,color}
\usepackage{latexsym,amsfonts}
\usepackage{amsmath}
\usepackage{amssymb}
\usepackage{bbm}

\usepackage{tikz}
\usepackage{url}
\usepackage{fullpage}

\definecolor{blue}{RGB}{0,82,147} 
\definecolor{red}{RGB}{202,033,063}

\colorlet{green}{green!50!black}
\usepackage[breaklinks=true]{hyperref} 
\hypersetup{
	colorlinks=true,
	citecolor=green,
	linkcolor=blue,
	urlcolor=black,
}
\newcounter{lpnumber} 
\newenvironment{linearprogram}[1]
{ \stepcounter{lpnumber}
  \begin{gather} #1 \tag{LP\arabic{lpnumber}} \\[-5ex] \notag
  \end{gather}
  \hspace{1.5cm} subject to \\[-3ex]
  \align }
{ \endalign }
\newcommand{\minimize}[1]{\text{minimize} \ #1}
\newcommand{\maximize}[1]{\text{maximize} \ #1}

\newcommand{\vote}{\mathsf{vote}}
\newcommand{\wt}{\mathsf{wt}}
\newcommand{\poly}{\mathsf{poly}}
\newcommand{\cs}{\mathsf{score}}
\newcommand{\wcs}{\mathsf{\mbox{\sf{wt-score}}}}
\newcommand{\wins}{\mathsf{wins}}
\newcommand{\losses}{\mathsf{loss}}
\newcommand{\ties}{\mathsf{ties}}

\newtheorem{new-claim}{Claim}
\newtheorem{obs}{Observation}

\usepackage[switch]{lineno}
\usepackage[linesnumbered,lined,boxed,ruled,vlined]{algorithm2e} 

\SetKwInput{Parameters}{Parameters}
\SetKwComment{Comment}{$\triangleright$\ }{}
\SetAlFnt{\small}
\SetAlCapFnt{\small}
\SetAlCapNameFnt{\small}
\SetAlCapHSkip{0pt}
\IncMargin{-\parindent}

\usepackage{tikz}
\usetikzlibrary{shapes,shapes.multipart,decorations.text,arrows,decorations.markings,decorations.pathmorphing,shapes.geometric,positioning,decorations.pathreplacing}
\tikzset{snake it/.style={decorate, decoration=snake}}
\usepackage{caption}
\usepackage{subcaption}
\usepackage{mathtools}
\usepackage{thmtools,thm-restate}

\newtheorem{theorem}{Theorem}[section]
\newtheorem{lemma}[theorem]{Lemma}
\newtheorem{definition}[theorem]{Definition}

\newtheorem{proposition}[theorem]{Proposition}
\newtheorem{corollary}[theorem]{Corollary}

\newtheorem{remark}[theorem]{Remark}

\definecolor{mygreen}{rgb}{0.1, 0.6, 0.1}

\title{Semi-Popular Matchings and Copeland Winners\thanks{A preliminary version of this paper appeared in AAMAS '23: Proceedings of the 2023 International Conference on Autonomous Agents and Multiagent Systems, pages~957–965.}}

\author{
    \begin{tabular}{m{0.16\textwidth}m{0.16\textwidth}m{0.16\textwidth}m{0.16\textwidth}m{0.16\textwidth}m{0.16\textwidth}}
        \multicolumn{3}{c}{\textbf{Telikepalli Kavitha}} & \multicolumn{3}{c}{\textbf{Rohit Vaish}}\\
		\multicolumn{3}{c}{\small{TIFR Mumbai}} & \multicolumn{3}{c}{\small{IIT Delhi}}\\
        \multicolumn{3}{c}{\href{mailto:kavitha@tifr.res.in}{\small{\texttt{kavitha@tifr.res.in}}}} & \multicolumn{3}{c}{\href{mailto:rvaish@iitd.ac.in}{\small{\texttt{rvaish@iitd.ac.in}}}}\\
    \end{tabular}
}

\newcommand{\BibTeX}{\rm B\kern-.05em{\sc i\kern-.025em b}\kern-.08em\TeX}

\begin{document}
\date{}
\maketitle

\begin{abstract}
Given a graph $G = (V,E)$ where every vertex has a weak ranking over its neighbors, we consider the problem of computing an {\em optimal} matching as per agent preferences. Classical notions of optimality such as {\em stability} and its relaxation {\em popularity} could fail to exist when $G$ is non-bipartite.
In light of the non-existence of a popular matching, we consider its relaxations that satisfy universal existence. We find a positive answer in the form of {\em semi-popularity}. A matching $M$ is semi-popular if for a majority of the matchings $N$ in $G$, $M$ does not lose a head-to-head election against $N$. We show that a semi-popular matching always exists in any graph~$G$ and complement this existence result with a fully polynomial-time randomized approximation scheme (FPRAS).

A special subclass of semi-popular matchings is the set of {\em Copeland winners}---the notion of Copeland winner 
is classical in social choice theory and a Copeland winner always exists in any voting instance. We study the complexity of computing a matching that is a Copeland winner and show there is no polynomial-time algorithm for this problem unless $\mathsf{P} = \mathsf{NP}$.
\end{abstract}

\section{Introduction}
\label{sec:intro}

Matching problems with preferences are of central importance in economics, computer science, and operations research~\cite{GI89,M13,RS92}. Over the years, these problems have found several real-world applications such as in school choice~\cite{APR05,APR+05}, labor markets~\cite{R84,RP99}, and dormitory assignment~\cite{PPR08}. The input in such problems is typically a graph $G = (V,E)$ where the vertices correspond to \emph{agents}, each with a weak ranking of its neighbors. The goal is to divide the agents into pairs, i.e., find a matching in~$G$, while optimizing some criterion of agent satisfaction based on their preferences.

A classical criterion of agent satisfaction in the matching literature is \emph{stability} which requires that there is no {\em blocking edge}, i.e., no pair of agents simultaneously prefer each other over their prescribed matches~\cite{GS62}. 
Stability is an intuitively appealing notion, but it can be too demanding in the context of general (i.e., not necessarily bipartite) graphs, also known as {\em roommates instances}. Indeed, there exist simple roommates instances that do not admit any stable matching~(Fig.~\ref{fig:Intro_Stable} and Fig.~\ref{fig:Intro_Popular_Roommates}). 

\begin{figure}[h]
\small
\captionsetup[subfigure]{justification=centering}
    \centering
    \begin{subfigure}[b]{0.35\linewidth}
        \centering
        \begin{tikzpicture}
            \tikzset{mynode/.style = {shape=circle,draw,fill=black,inner sep=1.5pt}}
            \tikzset{edge/.style = {solid}}
            \node[mynode] (1) at (0,0) {};
            \node (1a) at (-0.3,0) {$a$};
            \node[mynode] (2) at (3,0) {};
            \node (2a) at (3.3,0) {$b$};
            \node[mynode] (3) at (1.5,2.5) {};
            \node (3a) at (1.5,2.8) {$c$};
            \node[mynode] (4) at (1.5,1.1) {};
            \node (4a) at (1.5,0.8) {$d$};
            \draw[edge] (1) to node [near start,fill=white,inner sep=0pt] (121) {$1$} node [near end,fill=white,inner sep=0pt] (122) {$2$} (2);
            \draw[edge] (3) to node [near start,fill=white,inner sep=0pt] (321) {$2$} node [near end,fill=white,inner sep=0pt] (322) {$1$} (2);
            \draw[edge] (1) to node [near start,fill=white,inner sep=0pt] (123) {$2$} node [near end,fill=white,inner sep=0pt] (321) {$1$} (3);
            \draw[edge] (1) to node [near start,fill=white,inner sep=0pt] (124) {$3$} node [near end,fill=white,inner sep=0pt] (421) {$1$} (4);
            \draw[edge] (2) to node [near start,fill=white,inner sep=0pt] (224) {$3$} node [near end,fill=white,inner sep=0pt] (422) {$2$} (4);
            \draw[edge] (3) to node [near start,fill=white,inner sep=0pt] (324) {$3$} node [near end,fill=white,inner sep=0pt] (423) {$3$} (4);
        \end{tikzpicture}
        \subcaption{}
        \label{fig:Intro_Stable}
    \end{subfigure}
    \hfill
    \begin{subfigure}[b]{0.27\linewidth}
        \centering
        \begin{tikzpicture}
            \tikzset{mynode/.style = {shape=circle,draw,fill=black,inner sep=1.5pt}}
            \tikzset{edge/.style = {solid}}
            \node[mynode] (1) at (0,0) {};
            \node (1a) at (-0.3,0) {$a$};
            \node[mynode] (2) at (2.5,0) {};
            \node (2a) at (2.8,0) {$b$};
            \node[mynode] (3) at (1.25,1.5) {};
            \node (3a) at (1.25,1.8) {$c$};
            \draw[edge] (1) to node [near start,fill=white,inner sep=0pt] (121) {$1$} node [near end,fill=white,inner sep=0pt] (122) {$2$} (2);
            \draw[edge] (2) to node [near start,fill=white,inner sep=0pt] (231) {$1$} node [near end,fill=white,inner sep=0pt] (232) {$2$} (3);
            \draw[edge] (3) to node [near start,fill=white,inner sep=0pt] (311) {$1$} node [near end,fill=white,inner sep=0pt] (312) {$2$} (1);
        \end{tikzpicture}
        \subcaption{}
        \label{fig:Intro_Popular_Roommates}
    \end{subfigure}
    \hfill
    \begin{subfigure}[b]{0.35\linewidth}
        \centering
        \begin{tikzpicture}
            \tikzset{mynode/.style = {shape=circle,draw,fill=black,inner sep=1.5pt}}
            \tikzset{edge/.style = {solid}}
            \node[mynode] (1) at (0,0) {};
            \node[mynode] (2) at (2.5,0) {};
            \node[mynode] (3) at (0,1.2) {};
            \node[mynode] (4) at (2.5,1.2) {};
            \node[mynode] (5) at (0,2.4) {};
            \node[mynode] (6) at (2.5,2.4) {};
            \draw[edge] (1) to node [near start,fill=white,inner sep=0pt] (122) {$3$} node [near end,fill=white,inner sep=0pt] (221) {$1$} (2);
            \draw[edge] (1) to node [near start,fill=white,inner sep=0pt] (124) {$2$} node [near end,fill=white,inner sep=0pt] (421) {$1$} (4);
            \draw[edge] (1) to node [near start,fill=white,inner sep=0pt] (126) {$1$} node [near end,fill=white,inner sep=0pt] (621) {$1$} (6);
            \draw[edge] (3) to node [near start,fill=white,inner sep=0pt] (322) {$3$} node [near end,fill=white,inner sep=0pt] (223) {$1$} (2);
            \draw[edge] (3) to node [near start,fill=white,inner sep=0pt] (324) {$2$} node [near end,fill=white,inner sep=0pt] (423) {$1$} (4);
            \draw[edge] (3) to node [near start,fill=white,inner sep=0pt] (326) {$1$} node [near end,fill=white,inner sep=0pt] (623) {$1$} (6);
            \draw[edge] (5) to node [near start,fill=white,inner sep=0pt] (522) {$3$} node [near end,fill=white,inner sep=0pt] (225) {$1$} (2);
            \draw[edge] (5) to node [near start,fill=white,inner sep=0pt] (524) {$2$} node [near end,fill=white,inner sep=0pt] (425) {$1$} (4);
            \draw[edge] (5) to node [near start,fill=white,inner sep=0pt] (526) {$1$} node [near end,fill=white,inner sep=0pt] (625) {$1$} (6);
        \end{tikzpicture}
        \subcaption{}
        \label{fig:Intro_Popular_Ties}
    \end{subfigure}
    \caption{\small{In each figure, the vertices denote the agents, and the number closer to a vertex denotes its rank for the other vertex, where a lower number denotes a more preferred neighbor. (a) An instance with no stable matching but with two popular matchings $\{(a,d),(b,c)\}$ and $\{(a,c),(b,d)\}$. (b) A roommates instance without a popular matching. (c)~A bipartite instance (with ties) without a popular matching.}}
    \label{fig:Intro}
\end{figure}

Popularity is a meaningful relaxation of stability that captures welfare in a collective sense~\cite{G75}.
Intuitively, popularity asks for a matching that is not {\em defeated} by any matching in a head-to-head comparison. 
More concretely, consider an election in which the matchings play the role of candidates and the agents/vertices act as voters. Given a pair of matchings $M$ and $N$, a vertex prefers $M$ to $N$ (resp., $N$ to $M$) if it gets a more preferred partner in $M$ (resp.,~$N$).

In the $M$-vs-$N$ election, every vertex votes for the matching in $\{M,N\}$ that it prefers
and it abstains from voting if it is indifferent. Note that being left unmatched is the worst choice for any voter.
In this $M$-vs-$N$ election, let $\phi(M,N)$ be the number of votes for $M$ and let $\phi(N,M)$ be the number of votes for $N$. 
Further, let $\Delta(M,N) \coloneqq \phi(M,N) - \phi(N,M)$. We say that the matching $N$ \emph{defeats} (or is \emph{more popular} 
than) the matching $M$ if $\Delta(N,M)>0$. A \emph{popular} matching is one such that there is {\em no} ``more popular'' matching.

\begin{definition}[\textbf{Popular matching}]
\label{def:popular_Intro}
A matching $M$ is popular if there is no matching that is more popular than $M$, i.e., $\Delta(M,N) \ge0$ for all matchings $N$.
\end{definition}

Thus, a popular matching is a \emph{weak Condorcet winner} in the underlying 
election among matchings~\cite{C17}.\footnote{A matching $M$ is a \emph{Condorcet winner} if $\Delta(M,N) > 0$ for every matching $N \ne M$ in $G$, and a \emph{weak Condorcet winner} if $\Delta(M,N) \geq 0$ for every matching $N$ in $G$.\label{footnote:Condorcet}} 
Note that under strict preferences, a stable matching is also popular~\cite{C00}, but a popular matching can exist in instances with no stable matching; e.g., there are two popular matchings $\{(a,d),(b,c)\}$ and $\{(a,c),(b,d)\}$ in Fig.~\ref{fig:Intro_Stable}.
Thus popularity is a more relaxed criterion than stability---it ensures
``collective stability'' as there is no matching that makes more agents better off than those who are worse off.

Unfortunately, the popularity criterion also suffers from similar limitations as stability and more. First, although  popular matchings always exist in a bipartite graph with strict preferences~\cite{G75}, they could \emph{fail to exist} 
with weak rankings, i.e., when preferences include ties~(Fig.~\ref{fig:Intro_Popular_Ties}) 
or when the graph is non-bipartite~(Fig.~\ref{fig:Intro_Popular_Roommates}).  
Second, determining the existence of a popular matching is known to be $\mathsf{NP}$-hard in roommates instances with strict preferences~\cite{FKPZ19,GMSZ19} and in bipartite instances with weak rankings~\cite{BIM10,CHK15}.

\subsection{Relaxations of Popularity}

The non-existence and intractability results for popular matchings motivate the study of relaxations in search of positive results.
A natural relaxation of popularity is {\em low unpopularity} and two of the well-known measures for 
quantifying  the unpopularity of a matching are {\em unpopularity margin} and {\em unpopularity factor}~\cite{M08}. The former bounds the additive gap and the latter
bounds  the multiplicative gap of the worst pairwise defeat suffered by the matching. Specifically, a matching $N$ that minimizes $\max_M\Delta(M,N)$ is a least unpopularity margin matching and a matching $N$ that minimizes
$\max_M (\phi(M,N)/\phi(N,M)$) is a least unpopularity factor matching. A popular matching has unpopularity margin~0 and unpopularity factor exactly~1 (where we interpret $0/0$ as 1). 

A matching $M$ with low unpopularity margin/factor may lose many elections, however $M$ can be considered to be {\em approximately popular} 
because there are no heavy defeats. 
It is easy to construct a bipartite instance on $n$ vertices with weak rankings (analogous to the one in Fig.~\ref{fig:Intro_Popular_Ties}) 
where every matching has unpopularity margin/factor $\Omega(n)$. So it can be the case that every matching suffers a heavy defeat against some other matching.

An intriguing alternative is to ask for a matching that does not suffer many defeats. If $M$ is a popular matching, the constraints $\Delta(M,N) \ge 0$ have to be satisfied for all matchings $N$. Finding a matching that satisfies the maximum number of these constraints is $\mathsf{NP}$-hard since that would 
solve the popular matching problem. What would be the complexity of seeking a matching that violates only a small 
fraction of these constraints?

Note that there are four matchings in the instance in Fig.~\ref{fig:Intro_Popular_Roommates} (this includes the empty matching) and each of these four matchings loses to at least one matching. Is there always a matching that is guaranteed to {\em not lose} against a good fraction of matchings, say a majority of the
matchings (i.e., not lose against at least $\mu/2$ matchings, where $\mu$ is the total number of matchings)?
This is precisely the notion of {\em semi-popularity} that was introduced in~\cite{K20}.

\begin{definition}[\textbf{Semi-popular matching}]
  A matching $M$ is semi-popular if $\Delta(M,N)~\geq~0$ for a majority of matchings $N$ in $G$.
\end{definition}  

Thus, matching $M$ is {\em semi-popular} if $M$ is undefeated by a majority of matchings, i.e., $M$ loses to at most $\mu/2$ matchings, where $\mu$ is the total
number of matchings in $G$. Note that the three matchings $M_1 = \{(a,b)\}, M_2 = \{(b,c)\}$, and $M_3 = \{(c,a)\}$ in Fig.~\ref{fig:Intro_Popular_Roommates} are semi-popular.
Popularity relies on the notion of {\em majority}, i.e., there is no matching that is preferred to a popular matching by more than half the agents. Semi-popularity takes the notion of majority a step further by asking for a matching undefeated 
by a majority of matchings.

Regarding a matching undefeated by many matchings to be {\em approximately popular} is in the same spirit as regarding a matching that does not have
many blocking edges to be {\em approximately stable}. When stable matchings do not exist in a roommates instance $G = (V,E)$, the complexity of finding a matching with the smallest number of blocking edges was studied in \cite{ABM06}, where this problem was shown to be $\mathsf{NP}$-hard. Further, even under strict preferences, this problem cannot be approximated within a factor of $|V|^{1-\varepsilon}$ for any $\varepsilon > 0$, unless $\mathsf{P} = \mathsf{NP}$~\cite{ABM06}. 

Semi-popular matchings were introduced in \cite{K20} to design an efficient bicriteria approximation algorithm for the min-cost popular matching problem in a bipartite instance with strict preferences (this is an $\mathsf{NP}$-hard problem~\cite{FKPZ19}). Stable matchings, and therefore semi-popular matchings, always exist in such instances, but what about general instances? We consider the following questions here.

\begin{quote}
    \emph{Does a semi-popular matching always exist in any roommates instance with weak rankings? If so, is it easy to find one?}
\end{quote}

We show a positive answer to the first question above and an {\em almost} positive answer to the second question above.
Our first observation is the following.

\begin{restatable}{proposition}{SemiCopelandExistence}
  \label{thm:semi}
  Every roommates instance where agents have weak rankings admits a semi-popular matching. 
\end{restatable}

The proof of Proposition~\ref{thm:semi} uses an averaging argument over the space of all matchings and is non-constructive. Thus, the above existence result for semi-popular
matchings does not automatically provide an efficient algorithm for computing such a matching. 
So though we know that semi-popular matchings always exist, the complexity of finding one remains elusive. However we are able to show an efficient randomized algorithm that with high probability~(specifically, with probability at least $1 - 1/|V|$) finds an {\em almost} semi-popular matching.

\begin{restatable}[\textbf{FPRAS for a semi-popular matching}]{theorem}{SemiCopelandFPRAS}
\label{thm:FPRAS_semi}
Given a roommates instance $G = (V,E)$ with weak rankings and any $\varepsilon > 0$, we can compute in $\poly(|V|,1/\varepsilon)$
time a matching $M$ such that $\Delta(M,N) \ge 0$ for at least $1/2-\varepsilon$ fraction of all matchings $N$ in $G$  with 
high probability.
\end{restatable}

Though we do not know how to find a semi-popular matching in polynomial time, we can find in $\poly(|V|,1/\varepsilon)$ time 
a matching $M$ that is undefeated by at least $1/2-\varepsilon$ fraction of the matchings with high probability.
It is relevant to note that our algorithm works for \emph{any} input graph $G$ (not necessarily bipartite) and can also accommodate \emph{weak rankings} (i.e., preferences with ties) and, more generally, partial order preferences. Thus, the notion of semi-popularity satisfies universal existence (Proposition~\ref{thm:semi}) 
and there is an efficient algorithm for computing an arbitrarily close approximation to it~(Theorem~\ref{thm:FPRAS_semi}) in the general roommates model.

For any matching $M$ in the graph $G$, let $\wins(M)$ (resp., $\ties(M)$) be the number of matchings that are defeated by (resp., tie with) $M$ in their head-to-head election. Deciding if there exists a matching $M$ that satisfies $\wins(M) + \ties(M) = \mu$ (where $\mu$ is the total number of matchings) is  $\mathsf{NP}$-hard~\cite{BIM10} as this is precisely the popular matching problem.
In contrast to this, there is a polynomial
time algorithm~\cite{BB20} to decide whether there exists a matching $M$ such that $\wins(M) = \mu - 1$, (so $M$ defeats every
other matching). Such a matching $M$ is a {\em Condorcet winner}\footnote{Recall that a Condorcet winner is one that wins every head-to-head election (see footnote~\ref{footnote:Condorcet}).} (aka a {\em strongly popular} matching). Of course, such a matching need not
exist in the given instance and this motivates the following natural question:

\begin{quote}
    \emph{Is there an efficient algorithm to find a matching $M$ that maximizes $\wins(M)$?}
\end{quote}

\paragraph{\bf Copeland winners.} 
The above question is closely connected to a classical notion called {\em Copeland winner} in social choice theory.
The Copeland rule is a well-known {\em Condorcet-consistent} voting rule (i.e., it selects the Condorcet winner whenever one exists) that has a long history starting from the 13th century and is named after Arthur H. Copeland~\cite{C51,HP01}.  Copeland's method is a natural extension of the Condorcet method 
and has been called ``perhaps the simplest modification'' of the Condorcet method~\cite{DM04}. Variants of the Copeland rule are used in sports leagues around the world. Below we define this method in the setting of matchings.

The {\em Copeland score} of a matching $M$ is defined as $\cs(M) \coloneqq \wins(M) + \ties(M)/2$. That is, the Copeland rule assigns one point for every win, half a point for every tie (this includes comparing the matching against itself), and none for a loss in a head-to-head comparison.

\begin{definition}
\label{def:copeland-winner}
A matching with the maximum value of $\cs(\cdot)$ is a \emph{Copeland winner}.   
\end{definition}

Social choice theory tells us that a Copeland winner satisfies many standard desirable properties such as Condorcet-consistency, monotonicity, majority~\cite{BCE+16,wiki-copeland,Pacuit}, and most importantly, a Copeland winner always exists. It is easy
to show that every Copeland winner is semi-popular (see Corollary~\ref{cor1}). 
However, unlike semi-popular matchings where we do not know the complexity of computing an \emph{exact} solution, it can be shown that finding an exact Copeland winner is computationally intractable.

\begin{restatable}[\textbf{Hardness of Copeland winner}]{theorem}{Copeland}
\label{thm:Copeland_NPHard}
Unless $\mathsf{P} = \mathsf{NP}$, there is no polynomial-time algorithm for finding a Copeland winner in a roommates instance with weak rankings.
\end{restatable}

The class of Copeland$^\alpha$ rules generalizes the Copeland rule where wins/ties/losses get the weights of $1/\alpha/0$ for some $0 \le \alpha \le 1$~\cite{FHS08,VMAB16}. 
Let us call a matching with the maximum value of $\wins(M) + \alpha\cdot\ties(M)$ a Copeland$^{\alpha}$ winner. 
So a Copeland$^0$ winner is a matching $M$ that maximizes $\wins(M)$.

It is easy to extend our proof of Theorem~\ref{thm:Copeland_NPHard} to show that unless $\mathsf{P} = \mathsf{NP}$, there is 
no polynomial-time algorithm in a roommates instance with weak rankings to find a Copeland$^{\alpha}$ winner
for any $0 \le \alpha < 1$. 
Thus in spite of the tractability of testing if there exists a matching $M$ with $\wins(M) = \mu -1$ and finding one if
so~\cite{BB20}, our earlier question on the tractability of finding a matching $M$ that maximizes $\wins(M)$ has a negative answer. That is, under standard complexity-theoretic assumptions, there is no polynomial-time algorithm for finding a matching $M$ that maximizes $\wins(M)$.

\paragraph{\bf Background and related work.}
There are polynomial-time algorithms known for deciding if a roommates instance with strict preferences admits a stable matching~\cite{Irv85}. As mentioned earlier, it is $\mathsf{NP}$-hard to decide if a popular matching exists in bipartite graphs with weak rankings or in non-bipartite graphs with strict preferences~\cite{BIM10,CHK15,FKPZ19,GMSZ19}. See Fig.~\ref{fig:All_Notions} for an illustration of relaxations among the various notions mentioned here.

\begin{figure}[t]
     \centering
     \tikzset{every picture/.style={line width=1pt}}
     \begin{tikzpicture}
     \footnotesize
		\draw[fill=green!5] (0,0.2) rectangle (8,3);
		\node (1) at (4,2.7) {\normalsize{Semi-Popular}};
		\draw [dashed,fill=red!15,rotate around={15:(5.1,1.4)}]
		(5.1,1.4) ellipse (2.4 cm and 0.7 cm);
		\node (1) at (6.1,1.7) {\normalsize{Popular}};
		\draw [fill=red!15,rotate around={-15:(2.9,1.4)}]
		(2.9,1.4) ellipse (2.4 cm and 0.7 cm);
		\draw [dashed,fill=none,rotate around={15:(5.1,1.4)}]
		(5.1,1.4) ellipse (2.4 cm and 0.7 cm);
		\node (1) at (1.9,1.7) {\normalsize{Copeland}};
		\draw[dashed,fill=green!17] (4,1) ellipse (1 cm and 0.4 cm);
		\node (1) at (4,1) {\normalsize{Condorcet}};
\end{tikzpicture}
\caption{\small Relationship among the various notions mentioned in this paper for the setting of roommates instances with weak preferences. A solid (resp., dashed) border indicates that the property is guaranteed to exist (resp., could fail to exist). Computational tractability (resp., intractability) is indicated via {\color{mygreen} green} (resp., {\color{red} red}) color. We use a lighter shade of {\color{mygreen} green} for the outer box to denote tractability of the ``almost'' variant of the problem.}
\vspace{-0.15in}
\label{fig:All_Notions}
\end{figure}

Algorithmic aspects of popular matchings have been extensively studied in the last fifteen years within theoretical computer science and combinatorial optimization literature and we refer to \cite{C17} for a survey. The special case of popular matchings in bipartite graphs with strict preferences has been of particular interest, where such matchings always exist, and the work that is closest to ours here is \cite{K20}, where semi-popular matchings were introduced.

\subsection{Our Techniques}
We establish our algorithmic result (Theorem~\ref{thm:FPRAS_semi}) by using a sampling-based procedure (Algorithm~\ref{alg:FPRAS_Semi_Copeland}). 
Sampling matchings from a near-uniform distribution is well-studied in theoretical computer science~\cite{MR95},
however it has not really been explored much in computational social choice. We use the sampling approach to search for an almost semi-popular matching in the exponentially large space of all matchings in $G = (V,E)$.
Specifically, we draw two independent samples, each containing $\Theta(\log |V| / \varepsilon^2)$ matchings, from a distribution that is $\varepsilon/4$-close to the uniform 
distribution in total variation distance.\footnote{Informally, the total variation distance between two probability distributions is the largest possible difference between the probabilities that the two probability distributions can assign to the same event.} By the seminal result of Jerrum and Sinclair~\cite{JS89}, there is an algorithm with running time $\poly(|V|,\log( 1/\varepsilon))$ for generating a sample from such a distribution.

We then pit the two random samples against each other by evaluating all head-to-head elections between pairs of matchings, one matching from each sample, and pick the one with the highest Copeland score in these elections. It is easy to see that the chosen matching is semi-popular `on the sample'~(Lemma~\ref{lemma2}). By a standard concentration argument, we are able to show that this matching is almost semi-popular with respect to \emph{all} the matchings in the given instance with high probability~(Lemma~\ref{lem:correct}).

Our hardness result for Copeland winners (Theorem~\ref{thm:Copeland_NPHard}) uses a reduction from \textsf{VERTEX COVER}.\footnote{In the VERTEX COVER problem, the input is a graph $G = (V,E)$ and an integer $k$, and the goal is to determine if there is a subset $S \subseteq V$ of at most $k$ vertices such that every edge in $E$ is incident to some vertex in $S$. This is an NP-hard problem.} At a high level, our construction is inspired by a
construction in \cite{FK22} that used a far simpler instance to show that the extension complexity of the bipartite popular matching polytope is near-exponential. Our construction, on the other hand, is considerably more involved and we construct a non-bipartite instance $G$ with weak rankings. What makes our construction particularly tricky is that \emph{every} Copeland winner in $G$ has to correspond to a minimum vertex cover in the input instance~$H$. In general, we do not know how to characterize Copeland winners in $G$. In fact, we do not even know how to test if a given matching is a Copeland winner or not. This makes our reduction 
challenging. We use the LP framework for popular matchings to analyze Copeland winners and this leads to our hardness proof. 
This proof is given in Section~\ref{sec:hardness}. 

\section{Computing an Almost Semi-Popular Matching}
\label{sec:Results}

Our input is a roommates instance $G = (V,E)$ on $n$ vertices where every vertex has a weak ranking over its neighbors. While it is easy to construct roommates instances that admit no popular matchings (see Fig.~\ref{fig:Intro_Popular_Roommates}), semi-popular matchings are always present in any instance~$G$, as we show below.

Let $\mu$ be the total number of matchings in $G$.
For any matching $M$ in the graph $G$, recall that $\wins(M)$ (resp., $\ties(M)$) is the number of matchings that are defeated by (resp., tie with) $M$ in their head-to-head election. A matching $M$ is popular if and only if $\wins(M) + \ties(M) = \mu$ and $M$ is semi-popular if and only if $\wins(M) + \ties(M) \ge \mu/2$. Recall that 
$\cs(M) = \wins(M) + \ties(M)/2$. The following lemma immediately implies Proposition~\ref{thm:semi}.

\begin{lemma}
  \label{lem:revised1}
  Every roommates instance (where agents have weak rankings) admits a matching $M$ with $\cs(M) \ge \mu/2$.
\end{lemma}

Lemma~\ref{lem:revised1} is a straightforward consequence of the following observation which, in turn, follows easily from pigeonhole principle.\footnote{We thank an anonymous reviewer for suggesting to use this fact to prove Lemmas~\ref{lem:revised1} and~\ref{lemma2}.}

\begin{proposition}
  \label{prop:directed}
  In any directed multigraph with $n$ vertices and $m$ edges, there exists some vertex with outdegree at least $\lceil m/n \rceil$.
\end{proposition}

\begin{proof} (of Lemma~\ref{lem:revised1})
Construct a directed multigraph whose vertices correspond to the matchings. Between any pair of vertices $(u,v)$ in this multigraph, add \emph{two} directed edges from $u$ to $v$ if the matching corresponding to $u$ defeats the one corresponding to $v$. Otherwise, if the corresponding matchings are tied, add one directed edge from $u$ to $v$ and another from $v$ to $u$.

Finally, add a directed edge (self-loop) from every vertex to itself.

Observe that a vertex in the multigraph constructed above has outdegree $d$ if and only if the corresponding matching has score $d/2$. The multigraph has $\mu(\mu-1) + \mu$ edges. Thus, by Proposition~\ref{prop:directed}, there must exist a vertex with outdegree at least $\mu$. Then, the corresponding matching, say $M$, must have $\cs(M)$ at least $\mu/2$.
\end{proof}

\begin{corollary}
\label{cor1}
Every Copeland winner is a semi-popular matching.
\end{corollary}
\begin{proof}
Let $M$ be a Copeland winner. Then $\cs(M) \ge \mu/2$ (by Lemma~\ref{lem:revised1}). Since $\cs(M) = \wins(M) + \ties(M)/2 
\le \wins(M) + \ties(M)$, we have $\wins(M) + \ties(M) \ge \mu/2$. Thus $M$ is semi-popular.
\end{proof}

\noindent{\bf Our algorithm.}
We will now show an FPRAS for computing a semi-popular matching. In fact, we will construct a matching $M$ with $\cs(M) \ge (1 -\varepsilon) \cdot \mu/2$ with high probability. In order to construct such a matching, as mentioned earlier, we will use the classical result from~\cite{JS89} that shows an algorithm with running time $\poly(n, \log(1/{\varepsilon}))$ to sample matchings from a distribution $\varepsilon$-close to the uniform distribution in total variation distance~(see \cite[Corollary~4.3]{JS89}).

Our algorithm is presented as Algorithm~\ref{alg:FPRAS_Semi_Copeland}. The input to our algorithm is a roommates instance $G = (V,E)$ on $n$ vertices along with a parameter $\varepsilon > 0$. It computes two independent samples $\mathcal{S}_0$ and $\mathcal{S}_1$ of $k = \lceil (32 \ln n / \varepsilon^2) \rceil$  matchings---each from a distribution  $\varepsilon/4$-close to the uniform distribution (on all the matchings in $G$) in total variation distance.

\begin{algorithm}[t]
\DontPrintSemicolon
\footnotesize
 \linespread{1.1}
\KwIn{A graph $G = (V,E)$ on $n$ vertices and a set of weak rankings for every vertex $v \in V$.}
\Parameters{$\varepsilon > 0$.}
\KwOut{A matching in $G$.}

Produce two independent samples $\mathcal{S}_0$ and $\mathcal{S}_1$ of $k = \lceil (32 \ln n / \varepsilon^2) \rceil$ 
matchings where each matching is chosen from a distribution that is $\varepsilon/4$-close to the uniform distribution (on all matchings in $G$) in total variation distance.\label{algline:Sampling_Step}\;
\BlankLine
\ForEach{matching $M \in \mathcal{S}_0 \cup \mathcal{S}_1$}{
Initialize $\wins'_M = \ties'_M = 0$.\;
}
\ForEach{matching $M \in \mathcal{S}_0$}{
   \ForEach{matching $N \in \mathcal{S}_1$}{
	if $\Delta(M,N) > 0$ then $\wins'_M = \wins'_M + 1$.\;
	if $\Delta(M,N) = 0$ then $\ties'_M = \ties'_M + 1$ and $\ties'_N = \ties'_N + 1$.\;
    if $\Delta(M,N) < 0$ then $\wins'_N = \wins'_N + 1$.
	}
  }	
\KwRet{a matching $S \in \mathcal{S}_0 \cup \mathcal{S}_1$ with the maximum value of $\wins'_S + \ties'_S/2$.}
\caption{An FPRAS for Semi-Popular Matchings.}
\label{alg:FPRAS_Semi_Copeland}
\end{algorithm}

For $M \in \mathcal{S}_0$ (resp., $\mathcal{S}_1$), let $\wins'_M$ be the number of matchings in $\mathcal{S}_1$ (resp., $\mathcal{S}_0$) that $M$ wins against
and let $\ties'_M$  be the number of matchings in $\mathcal{S}_1$ (resp., $\mathcal{S}_0$) that $M$ ties with. 

Our algorithm computes $\cs'_S = \wins'_S + \ties'_S/2$ for each $S \in \mathcal{S}_0 \cup \mathcal{S}_1$.
It returns a matching in $\mathcal{S}_0 \cup \mathcal{S}_1$ with the maximum value of $\cs'$.
We will now show that such a matching 
has a high Copeland score on the sample. Recall that $|\mathcal{S}_0| = |\mathcal{S}_1| =  k$.

\begin{lemma}
\label{lemma2}
  If $S^*$ is the matching returned by Algorithm~\ref{alg:FPRAS_Semi_Copeland}, then $\cs'_{S^*} \ge k/2$.
\end{lemma}
\begin{proof}
    Consider a bipartite graph whose left and right vertex sets correspond to the matchings in $\mathcal{S}_0$ and $\mathcal{S}_1$, respectively; thus, there are $2k$ vertices overall. If the $i^\text{th}$ matching in $\mathcal{S}_0$ defeats (resp., is defeated by) the $j^\text{th}$ matching in $\mathcal{S}_1$, then add two directed edges from the $i^\text{th}$ left vertex to the $j^\text{th}$ right vertex (resp., from the $j^\text{th}$ right vertex to the $i^\text{th}$ left vertex). If the two matchings are tied, add two directed edges---one in either direction---between the corresponding vertices.

    Observe that a vertex in the bipartite graph constructed above has outdegree $d$ if and only if the corresponding matching, say $M$, has $\cs'_{M} \geq d/2$. There are $2k^2$ edges in the graph. Thus, by Proposition~\ref{prop:directed}, there must exist a vertex with outdegree at least $k$, and therefore a matching, say $M$, with $\cs'_{M} \geq k/2$. Thus, the matching $S^*$ returned by Algorithm~\ref{alg:FPRAS_Semi_Copeland} has  $\cs'_{S^*} \geq \cs'_{M} \geq k/2$.
\end{proof}

We will show that the on-sample guarantee of Lemma~\ref{lemma2} carries over, with high probability, to the entire set of matchings.
This proof makes use of a tail bound for the random variable $\cs'_S$ corresponding to the on-sample Copeland score of any fixed matching $S \in \mathcal{S}_0 \cup \mathcal{S}_1$. 

\begin{lemma}
  \label{lemma0}
  Let $S \in \mathcal{S}_0 \cup \mathcal{S}_1$ be any fixed matching sampled by Algorithm~\ref{alg:FPRAS_Semi_Copeland}. Then the probability that $\cs'_S \geq k\cdot(\cs(S)/\mu + \varepsilon/2)$ is at most $1/n$.
\end{lemma}  
\begin{proof}
Assume without loss of generality that $S \in \mathcal{S}_1$. If $\mathcal{S}_0$ was a set of $k$ matchings chosen uniformly at random from the set of all matchings in $G$, then the probability that $S$ defeats any matching in $\mathcal{S}_0$ is $\wins(S)/\mu$ and the probability that it ties with any matching in $\mathcal{S}_0$ is $\ties(S)/\mu$. 
 
 However, the matchings in $\mathcal{S}_0$ are sampled from a distribution $\varepsilon/4$-close to the uniform distribution in total variation distance. So, the probability that $S$ defeats any matching in $\mathcal{S}_0$ is at most $\wins(S)/\mu + \varepsilon/4$ and the probability that it ties with any matching in $\mathcal{S}_0$ is at most 
 $\ties(S)/\mu + \varepsilon/4$. 
 
 Observe that $\cs'_S = X_1 + \ldots + X_k$, where, for each $i$, the random variable 
$X_i \in \{0,\frac{1}{2},1\}$ denotes whether $S$ loses/ties/wins against the $i$-th matching in $\mathcal{S}_0$. Note that $\mathbb{E}[X_i] \le \wins(S)/\mu + \ties(S)/2\mu + 3\varepsilon/8$ for each $i$. Since $\cs(S) = \wins(S) + \ties(S)/2$, linearity of expectation gives
  \begin{equation}
    \label{eq1}
    \mathbb{E}\big{[}\cs'_S\big{]} \ \le \ k\cdot\left(\frac{\cs(S)}{\mu} + \frac{3\varepsilon}{8}\right).
  \end{equation}  

In light of Equation~\eqref{eq1}, to prove the lemma it suffices to bound the probability of the event that 
$$\cs'_S - \mathbb{E}[\cs'_S] \geq k\cdot\varepsilon/8,$$
\text{ or, equivalently, }
$$\cs'_S/k - \mathbb{E}[\cs'_S/k] \geq \varepsilon/8.$$ For this, we will use Hoeffding's inequality~\cite{H63}. Recall that if $X_1,\dots,X_k$ are bounded independent random variables such that $X_i \in [0,1]$ for all $i \in [k]$ and $Y \coloneqq (X_1 + \dots + X_k)/k$, then Hoeffding's inequality says that for any $t \geq 0$,
  \begin{equation*}
    \Pr\big{[}Y - \mathbb{E}[Y] \ \geq \ t\big{]} \ \le\ e^{-2kt^2}.
  \end{equation*}
Applying the above inequality for $Y \coloneqq \cs'_S/k$ and $t \coloneqq \varepsilon/8$, we get that
\begin{equation}
    \Pr\big{[}\cs'_S/k - \mathbb{E}[\cs'_S/k] \ \ge \ \varepsilon/8\big{]} \le\ e^{-k\varepsilon^2/32}.
    \label{eqn:Applying_Hoeffding}
\end{equation}
Substituting $k = \lceil (32 \ln n / \varepsilon^2) \rceil$ in Equation~\eqref{eqn:Applying_Hoeffding}, we get that the right-hand side is at most $1/n$. Thus, with probability at least $1 - 1/n$, we have $\cs'_S < \mathbb{E}[\cs'_S] + k\cdot\varepsilon/8$. Using the upper bound in Equation~\eqref{eq1}, it follows that $\cs'_S < k\cdot(\cs(S)/\mu + \varepsilon/2)$ with probability at least $1 - 1/n$.
\end{proof}

By Lemmas~\ref{lemma2} and \ref{lemma0}, the matching $S^*$ returned by our algorithm satisfies
$k/2 \ \le \ \cs'_{S^*} \, < \ k\cdot(\cs(S^*)/\mu + \varepsilon/2)$ with high probability.
Thus Lemma~\ref{lem:correct} follows.

\begin{lemma}
  \label{lem:correct}
  If $S^*$ is the matching returned by Algorithm~\ref{alg:FPRAS_Semi_Copeland}, then $\cs(S^*) > (1-\varepsilon) \cdot \mu/2$ with high probability.
\end{lemma}

So our algorithm computes a matching whose Copeland score is more than $(1-\varepsilon) \cdot \mu/2$ with high probability. Its running time is polynomial in $n$ and $1/\varepsilon$. Since  $\wins(S^*) + \ties(S^*) \ge \cs(S^*)$, Theorem~\ref{thm:FPRAS_semi} follows.
\SemiCopelandFPRAS*

\section{Finding a Copeland Winner: A Hardness Result}
\label{sec:hardness}
In Section~\ref{sec:outline} we will first give a high-level overview of the proof of Theorem~\ref{thm:Copeland_NPHard} which states that under standard complexity-theoretic assumptions, there is no polynomial-time algorithm for finding a Copeland winner. Several details are given in Section~\ref{sec:proof-Thm3} and the remaining details can be found in the Appendix.

\subsection{A High-Level Overview}
\label{sec:outline}
Given a \textsf{VERTEX COVER} instance 
$H = (V_H,E_H)$, we will construct a roommates instance $G = (V,E)$ such that any Copeland winner in $G$ will correspond to a minimum vertex cover in 
$H$. We will assume that the vertices in the \textsf{VERTEX COVER} instance are indexed as $1,2,\dots,n$, i.e., $V_H = \{1,\ldots,n\}$. Specifically,
\begin{itemize}
\item for every vertex $i \in V_H$, there is a gadget $Z_i$ in $G$ on 4 {\em main} vertices $a_i,b_i,a'_i,b'_i$ and 100
  {\em auxiliary} vertices $u_i^0,\ldots,u_i^{99}$, and
\item for every edge $e \in E_H$, there is a gadget $Y_e$ in $G$ on 6 {\em main} vertices
  $s_e,t_e,s'_e,t'_e,s''_e,t''_e$ along with 8 {\em auxiliary} vertices $v_e,v'_e,w_e,w'_e,c_e,d_e,c'_e,d'_e$ (see Fig.~\ref{fig:hardness2}).
\end{itemize}

\begin{figure*}[t]
\small
     \centering
     \begin{tikzpicture}
        \tikzset{mynode/.style = {shape=circle,draw,fill=black,inner sep=1.5pt}}
        \tikzset{solidedge/.style = {solid}}
        \tikzset{dashededge/.style = {dashed}}
%
\fill [green!8] (-0.7,-0.5) rectangle (3.2,5);
%
\fill [green!8] (4,-0.5) rectangle (10.5,5);
\node[mynode] (1) at (0,0) {};
\node (1a) at (-0.3,0) {$a'_i$};
\node[mynode] (2) at (2.5,0) {};
\node (2a) at (2.8,0) {$b'_i$};
\node[mynode] (3) at (0,1.5) {};
\node (3a) at (-0.3,1.5) {$a_i$};
\node[mynode] (4) at (2.5,1.5) {};
\node (4a) at (2.5,1.2) {$b_i$};
\node[mynode] (5) at (0,3) {};
\node (5a) at (-0.3,3) {$u_i^0$};
\node[mynode] (6) at (2.5,3) {};
\node (6a) at (2.9,3) {$u_i^{99}$};
\node[mynode] (7) at (0.5,3.5) {};
\node[mynode] (8) at (2,3.5) {};
\node[mynode] (9) at (1,4) {};
\node[mynode] (10) at (1.5,4) {};
\node[mynode] (11) at (6,0) {};
\node (11a) at (5.7,0) {$s''_e$};
\node[mynode] (12) at (8.5,0) {};
\node (12a) at (8.8,0) {$t''_e$};
\node[mynode] (13) at (6,1.5) {};
\node (13a) at (5.7,1.5) {$s'_e$};
\node[mynode] (14) at (8.5,1.5) {};
\node (14a) at (8.8,1.5) {$t'_e$};
\node[mynode] (15) at (6,3) {};
\node (15a) at (6,2.7) {$s_e$};
\node[mynode] (16) at (8.5,3) {};
\node (16a) at (8.5,2.7) {$t_e$};
\node[mynode] (17) at (6.4,4.2) {};
\node (17a) at (6.4,4.5) {$v_e$};
\node[mynode] (18) at (8.1,4.2) {};
\node (18a) at (8.1,4.5) {$w'_e$};
\node[mynode] (19) at (5.2,3.7) {};
\node (19a) at (5,4) {$v'_e$};
\node[mynode] (20) at (9.3,3.7) {};
\node (20a) at (9.5,4) {$w_e$};
\node[mynode] (21) at (4.8,2.6) {};
\node (21a) at (4.5,2.6) {$c_e$};
\node[mynode] (22) at (4.2,1.5) {};
\node (22a) at (4.2,1.2) {$d_e$};
\node[mynode] (23) at (9.7,2.6) {};
\node (23a) at (10,2.6) {$c'_e$};
\node[mynode] (24) at (10.3,1.5) {};
\node (24a) at (10.3,1.2) {$d'_e$};
\draw[solidedge,red] (1) to node [near start,fill=white,inner sep=0pt] (122) {$2$} node [near end,fill=white,inner sep=0pt] (221) {$2$} (2);
\draw[solidedge,red] (3) to node [near start,fill=white,inner sep=0pt] (324) {$1$} node [near end,fill=white,inner sep=0pt] (423) {$1$} (4);
\draw[solidedge,blue] (1) to node [near start,fill=white,inner sep=0pt] (124) {$1$} node [near end,fill=white,inner sep=0pt] (421) {$2$} (4);
\draw[solidedge,blue] (3) to node [near start,fill=white,inner sep=0pt] (322) {$2$} node [near end,fill=white,inner sep=0pt] (223) {$1$} (2);
\draw[solidedge] (3) to (5);
\draw[solidedge] (3) to (7);
\draw[solidedge] (3) to (9);
\draw[solidedge] (3) to (10);
\draw[solidedge] (3) to (8);
\draw[solidedge] (3) to (6);
\draw[solidedge,teal] (11) to node [very near start,fill=white,inner sep=0pt] (11212) {$1$} node [very near end,fill=white,inner sep=0pt] (12211) {$1$} (12);
\draw[solidedge,olive] (13) to node [very near start,fill=white,inner sep=0pt] (13214) {$1$} node [very near end,fill=white,inner sep=0pt] (14213) {$1$} (14);
\draw[solidedge,olive] (11) to node [very near start,fill=white,inner sep=0pt] (11216) {$2$} node [very near end,fill=white,inner sep=0pt] (16211) {$1$} (16);
\draw[solidedge,olive] (15) to node [very near start,fill=white,inner sep=0pt] (15212) {$3$} node [very near end,fill=white,inner sep=0pt] (12215) {$2$} (12);
\draw[solidedge,teal] (13) to node [near start,fill=white,inner sep=0pt] (13216) {$2$} node [near end,fill=white,inner sep=0pt] (16213) {$3$} (16);
\draw[solidedge,teal] (15) to node [near start,fill=white,inner sep=0pt] (15214) {$1$} node [near end,fill=white,inner sep=0pt] (14215) {$2$} (14);
\draw[solidedge] (15) to node [near start,fill=white,inner sep=0pt] (15217) {$5$} node [near end,fill=white,inner sep=0pt] (17215) {$1$} (17);
\draw[solidedge] (17) to node [near start,fill=white,inner sep=0pt] (17219) {$2$} node [near end,fill=white,inner sep=0pt] (19217) {$1$} (19);
\draw[solidedge] (15) to node [near start,fill=white,inner sep=0pt] (15219) {$4$} node [near end,fill=white,inner sep=0pt] (19215) {$2$} (19);
\draw[solidedge] (16) to node [near start,fill=white,inner sep=0pt] (16220) {$5$} node [near end,fill=white,inner sep=0pt] (20216) {$1$} (20);
\draw[solidedge] (16) to node [near start,fill=white,inner sep=0pt] (16218) {$4$} node [near end,fill=white,inner sep=0pt] (18216) {$2$} (18);
\draw[solidedge] (18) to node [near start,fill=white,inner sep=0pt] (18220) {$1$} node [near end,fill=white,inner sep=0pt] (20218) {$2$} (20);
\draw[solidedge] (15) to node [near start,fill=white,inner sep=0pt] (15221) {$2$} node [near end,fill=white,inner sep=0pt] (21215) {$2$} (21);
\draw[solidedge] (16) to node [near start,fill=white,inner sep=0pt] (16223) {$2$} node [near end,fill=white,inner sep=0pt] (23216) {$2$} (23);
\draw[solidedge] (21) to node [near start,fill=white,inner sep=0pt] (21222) {$1$} (22);
\draw[solidedge] (23) to node [near start,fill=white,inner sep=0pt] (23224) {$1$} (24);
\draw[solidedge] (4) to (22);
%
\end{tikzpicture}
\caption{\small (Left) The vertex gadget $Z_i$. The {\color{red} red} colored edges within $Z_i$ indicate a stable matching. (Right) The edge gadget $Y_e$ on 14 vertices. All unnumbered edges incident to a vertex should be interpreted as ``tied for the last acceptable position''. The edges $(c_e,d_e),(v_e,v'_e),(w_e,w'_e),(c'_e,d'_e)$ along with the {\color{olive}olive}-colored (resp., {\color{teal}teal}-colored) edges define the matching $F_e$ (resp., $L_e$). Here $e = (i,j)$ and 
$(b_i,d_e)$ is an inter-gadget edge.}
\label{fig:hardness2}
\end{figure*}

\smallskip
\noindent{\bf The gadgets.}
The preferences of the vertices in the vertex gadget $Z_i$ and the edge gadget $Y_e$ are shown in Fig.~\ref{fig:hardness2}. Observe that in the vertex gadget $Z_i$, all the vertices $u^0_i,\ldots,u^{99}_i$ are tied at the third position (which is the last acceptable position) in $a_i$'s preference order. Similarly, in the edge gadget $Y_e$ (where $e = (i,j)$), 
the vertices $b_i$ and $c_e$ are tied in $d_e$'s preference list and the vertices $b_j$ and $c'_e$ are tied in $d'_e$'s preference list.

\medskip

\noindent{\em {\color{red} Red} state vs {\color{blue} blue} state.}
Let $M$ be a matching in $G$. We will say a vertex gadget $Z_i$ is in {\color{red} red} state in $M$ if $\{(a_i,b_i),(a'_i,b'_i)\} \subset M$ and in
{\color{blue} blue} state if $\{(a_i,b'_i),(a'_i,b_i)\} \subset M$. 

\medskip

\noindent{\bf A high-level overview of our hardness reduction.} We will show that any Copeland winner $M$ in $G$ has the following properties.

\begin{itemize}
    \item $M$ does not use any {\em inter-gadget} edge, i.e., a shared edge between a vertex gadget and an edge gadget.
    \item For any vertex $i \in V_H$, its vertex gadget $Z_i$ is either in {\color{red} red} state or in {\color{blue} blue} state in $M$.
    \item For any edge $e = (i,j)$, at least one of $Z_i, Z_j$ has to be in {\color{blue} blue} state in $M$.
    \item The vertices $i$ with vertex gadgets $Z_i$ in {\color{blue} blue} state in $M$ form a {\em minimum vertex cover} in $H$.
\end{itemize}

The above properties will imply Theorem~\ref{thm:Copeland_NPHard}. For the sake of readability, we do not include
in Section~\ref{sec:proof-Thm3} the proofs of all the lemmas that together imply Theorem~\ref{thm:Copeland_NPHard};
the proofs of lemmas marked by an asterisk ($\star$) are given in the appendix. The hardness of finding a Copeland winner will follow from Lemmas~\ref{lem:minimum} and \ref{lem:structure2}
which are proved in Section~\ref{sec:proof-Thm3}. 

\Copeland*

\subsection{Proof of Theorem~\ref{thm:Copeland_NPHard}}
\label{sec:proof-Thm3}
Let $e \in E$. In the edge gadget $Y_e$ (see Fig.~\ref{fig:hardness2}), we will find it convenient to define the matchings 
$$F_e = \{(s_e,t''_e), (s'_e,t'_e), (s''_e,t_e), (v_e,v'_e), (w_e,w'_e), (c_e,d_e), (c'_e,d'_e)\}$$ and 
$$L_e = \{(s_e,t'_e), (s'_e,t_e), (s''_e,t''_e), (v_e,v'_e), (w_e,w'_e), (c_e,d_e), (c'_e,d'_e)\}.$$
These are highlighted with {\color{olive}olive} and {\color{teal}teal} colors in Fig.~\ref{fig:hardness2}, respectively.
The following two lemmas will be very useful to us. 
Lemma~\ref{lem:10} discusses the number of matchings that tie with the
matching $F_e$ (or $L_e$) in the subgraph restricted to the edge gadget $Y_e$ (see Fig.~\ref{fig:hardness2}).

\begin{restatable}[$\star$]{lemma}{lemmaFe}
\label{lem:10}
In the subgraph restricted to $Y_e$, there are exactly 10 matchings that are tied with $F_e$ and no matching defeats $F_e$. Furthermore, an analogous statement holds for the matching $L_e$.
\end{restatable}

Next, Lemma~\ref{lem:ties} shows that there is {\em no} matching within the subgraph restricted to $Y_e$ that ``does better'' than $F_e$ or $L_e$ in terms of the number
of matchings that defeat or tie with it.

\begin{restatable}[$\star$]{lemma}{lemmaTe}
  \label{lem:ties}
For any matching $T_e$ in the subgraph restricted to $Y_e$, there are at least 10 matchings within this subgraph that either defeat or tie with $T_e$.
\end{restatable}

Recall the {\color{red} red}/{\color{blue} blue} states of a vertex gadget described in Section~\ref{sec:hardness}.  Lemma~\ref{lem:structure1} stated below
shows that a Copeland winner matching that does not use any inter-gadget edge must have each vertex gadget in either {\color{red}red} or {\color{blue}blue} state.

\begin{restatable}[$\star$]{lemma}{lemStructure}
\label{lem:structure1}
Let $M$ be a Copeland winner in $G$. If $M$ does not use any inter-gadget edge, then in any 
vertex gadget $Z_i$, 
either $\{(a_i,b_i),(a'_i,b'_i)\} \subset M$ or $\{(a_i,b'_i),(a'_i,b_i)\} \subset M$.
\end{restatable}

\begin{obs}
Consider the subgraph induced on $Z_i$.
\begin{itemize}
\item The {\color{red} red} matching $R_i = \{(a_i,b_i),(a'_i,b'_i)\}$ is tied with 2 matchings in this subgraph.
These are $R_i$ itself and the {\color{blue} blue} matching $B_i = \{(a_i,b'_i),(a'_i,b_i)\}$.
\item The {\color{blue} blue} matching $B_i = \{(a_i,b'_i),(a'_i,b_i)\}$ is tied with 3 matchings in this subgraph: these are $B_i$ itself, the {\color{red} red} matching $R_i$, and the matching $\{(a_i,b_i)\}$. 
\end{itemize}
Moreover, no matching in this subgraph defeats either the {\color{red} red} matching $R_i$ or
the {\color{blue} blue} matching $B_i$.
\label{obs:Vertex_Gadget_Ties}
\end{obs}


It is straightforward to verify the above observation.
For any gadget $X$, let $M \cap X$ denote the edges of matching $M$ in the subgraph restricted to $X$.

\begin{restatable}[$\star$]{lemma}{lemmaBlue}
  \label{lem:blue}
  Let $e = (i,j) \in E$. Let $M$ be any matching in $G$ such that both $Z_i$ and $Z_j$ are in {\color{red} red} state in $M$.
  Then there are at least 100 matchings within $Y_e \cup Z_i \cup Z_j$ that defeat or tie with $M \cap (Y_e \cup Z_i \cup Z_j)$.
\end{restatable}

Next, we will show in Lemma~\ref{lem:vertex-cover} that in a Copeland winner matching that does not use any inter-gadget edge, for any edge gadget, at least one of its adjacent vertex gadgets must be in {\color{blue} blue} state. The proof of Lemma~\ref{lem:vertex-cover} will make use of the following key technical lemma. 

\begin{restatable}[$\star$]{lemma}{lemmaProperties}
Let $M^*$ be any matching in $G$ that satisfies the following three conditions:
\begin{enumerate}
    \item Every vertex gadget is either in {\color{red} red} or {\color{blue} blue} state.
    \item For every edge gadget, at least one of its adjacent vertex gadgets is in {\color{blue} blue} state.
    \item For each edge $e = (i,j)$ where $i < j$, if the vertex gadget $Z_i$ is in {\color{blue} blue} state, then $M^* \cap Y_e = F_e$ otherwise $M^* \cap Y_e = L_e$.
\end{enumerate}
Then (i) $M^*$ is popular in $G$ and (ii) any matching that contains an inter-gadget edge {\em loses} to $M^*$.
\label{lem:Properties}
\end{restatable}

We showed in Lemma~\ref{lem:structure1} that the first property stated in Lemma~\ref{lem:Properties} is obeyed by any Copeland winner that does not use any
inter-gadget edge. Lemma~\ref{lem:vertex-cover} stated below will show that the second property stated in Lemma~\ref{lem:Properties} is also obeyed by any Copeland 
winner that does not use any inter-gadget edge.

\begin{restatable}{lemma}{ZiORZj}
  \label{lem:vertex-cover}
  Let $M$ be a Copeland winner in $G$. If $M$ does not use any \emph{inter-gadget} edge, then, for every edge $e = (i,j)$, at least one of $Z_i,Z_j$ has to be in {\color{blue} blue} state in $M$.
\end{restatable}
\begin{proof}
Suppose, for contradiction, that both $Z_i$ and $Z_j$ are in {\color{red} red} state in $M$ for some edge $e = (i,j)$. Let $\mathcal{X}$ denote the set of all vertex and edge gadgets in the matching instance. Consider a partitioning of $\mathcal{X}$ into \emph{single} gadgets and \emph{auxiliary} gadgets. Each auxiliary gadget is a triple of an edge and two adjacent vertex gadgets $(Y_e,Z_i,Z_j)$ where $e = (i,j)$ is an edge such that both $Z_i$ and $Z_j$ are in {\color{red} red} state in $M$. 
While there is such an edge $e$ with both its vertex gadgets in {\color{red} red} state and these vertex gadgets are ``unclaimed'' by any other edge, the edge $e$ claims both its vertex gadgets and makes an auxiliary gadget out of these three gadgets. All the remaining vertex and edge gadgets are classified as single gadgets, including those edges one or both of whose endpoints have been claimed by some other edge(s). Observe that 
$$\losses(M) + \ties(M) \ge \Pi_{X\in \mathcal{X}}(\losses(M \cap X) + \ties(M \cap X))$$ where $X$ is any single or auxiliary gadget, and $\losses(M\cap X)$ (resp., $\ties(M\cap X)$) is the number of matchings that defeat (resp., tie with) $M\cap X$ within the gadget $X$. Also, $\losses(M)$ (resp., $\ties(M)$) is the number of matchings that defeat (resp., tie with) the matching $M$.

Consider any auxiliary gadget $X = (Y_e,Z_i,Z_j)$. We know that $\losses(M\cap X) + \ties(M\cap X)$ is at least 100 (by Lemma~\ref{lem:blue}). Thus, we have
\begin{equation}
  \label{eqn1}
  \losses(M) + \ties(M) \ \ge\ 2^{n'}\cdot 3^{n''}\cdot 10^{m'}\cdot 100^t,
\end{equation}  
where $n'$ (resp., $n''$) is the number of vertices present as single gadgets in {\color{red} red} (resp., {\color{blue} blue)} state, $m'$ is the number of edges that are present as single gadgets, and $t$ is the number of
auxiliary gadgets in the aforementioned partition. Note that we used Lemma~\ref{lem:ties} here in bounding $\losses(M \cap Y_e) + \ties(M \cap Y_e)$ by 10 for 
every edge gadget $Y_e$ that is present as a single gadget, and used Lemma~\ref{lem:structure1} and Observation~\ref{obs:Vertex_Gadget_Ties} in bounding 
$\losses(M \cap Z_i) + \ties(M \cap Z_i)$ by 2 (resp., 3) for 
every vertex gadget $Z_i$ (acting as single gadget) that is in {\color{red} red} (resp., {\color{blue} blue)} state.

We will now construct another matching with Copeland score higher than that of $M$ to establish the desired contradiction. Let $M^*$ be the matching obtained by converting both $Z_i$ and $Z_j$ in each auxiliary gadget $(Y_e,Z_i,Z_j)$ in our partition above into
{\color{blue} blue} state and let $M^*\cap Y_e$ be either $F_e$ or $L_e$~(it does not matter which). For every edge $e' = (i',j')$
such that $Y_{e'}$ is present as a single gadget, observe that at least one of $i',j'$ is either in {\color{blue} blue} state or in an auxiliary gadget. If it is $\min\{i',j'\}$ that is in {\color{blue} blue} state/auxiliary gadget, then let $M^*\cap Y_{e'} = F_{e'}$, else let $M^*\cap Y_{e'} = L_{e'}$. Any vertex gadget that is present as a single gadget remains in its original {\color{red} red} or {\color{blue} blue} state. 

Notice that $M^*$ satisfies the conditions in Lemma~\ref{lem:Properties}. So $M^*$ is popular in $G$, i.e., $\losses(M^*)=0$, and any matching that ties with $M^*$ must not use any inter-gadget edge. Therefore, the number of matchings that tie with $M^*$ across the entire graph $G$ is simply the product of matchings that tie with it on individual single and auxiliary gadgets. Hence, 

\begin{equation}
\label{eqn2}  
  \losses(M^*) + \ties(M^*) \ = \ \ties(M^*) \ = \ 2^{n'}\cdot 3^{n''} \cdot 10^{m'}\cdot 90^t,
\end{equation}  
where $n',n'',m'$, and $t$ are defined as in \eqref{eqn1}. Note that the bound of $90$ for an auxiliary gadget in (\ref{eqn2}) follows from taking the product of 3, 3, and 10, which is the number of matchings that tie with $M^*$ in the two vertex gadgets and their common edge gadget (see Observation~\ref{obs:Vertex_Gadget_Ties} and Lemma~\ref{lem:10}).

Recall that $\cs(N) = \wins(N) + \ties(N)/2 = \mu  - \losses(N) - \ties(N)/2$ for any matching $N$.
Comparing \eqref{eqn1} and \eqref{eqn2} along with the fact that $\losses(M) \ge 0 = \losses(M^*)$, we have
$\cs(M^*) > \cs(M)$, as long as there is even a single edge $(i,j)$ such that both $Z_i$ and $Z_j$ are in {\color{red} red} state in $M$. Indeed, 
$$\losses(M^*) + \ties(M^*)/2 \ = \ 2^{n'-1}\cdot 3^{n''} \cdot 10^{m'}\cdot 90^t, \text{ and }$$
\begin{align*}
    \losses(M) + \ties(M)/2 \ & \ge\ (\left(\losses(M) + \ties(M)\right)/2 \\
    & \ge\ 2^{n'-1}\cdot 3^{n''}\cdot 10^{m'}\cdot 100^t.
\end{align*}

This contradicts the fact that $M$ is a Copeland winner. Hence, for every edge $e = (i,j)$, at least one of $Z_i,Z_j$ must be in {\color{blue} blue} state in $M$. This proves Lemma~\ref{lem:vertex-cover}.
\end{proof}

In Lemma~\ref{lem:vertex-cover}, we showed that for any Copeland winner $M$ that does not use any inter-gadget edge, the vertices whose gadgets are in {\color{blue} blue} state in $M$ constitute a vertex cover in $H$. The next result 
shows that the set of such vertices is, in fact, a {\em minimum} vertex cover.

\begin{restatable}{lemma}{BlueDenotesVertexCover}
\label{lem:minimum}
Let $M$ be a Copeland winner in $G$. If $M$ does not use any \emph{inter-gadget} edge, then the vertices whose gadgets are in {\color{blue} blue} state in $M$ constitute a minimum vertex cover in~$H$.
\end{restatable}
\begin{proof}
 We know from Lemma~\ref{lem:vertex-cover} that $\losses(M) + \ties(M) \ge \Pi_S(\losses(M \cap S) + \ties(M \cap S))$ where 
 the product is over all gadgets $S$. Further, from Lemma~\ref{lem:structure1}, Lemma~\ref{lem:ties}, and Observation~\ref{obs:Vertex_Gadget_Ties}, we know that the right hand side in the above inequality is at least 
  $2^{n-k}\cdot3^k\cdot10^m$, where $k$ is the number of vertex gadgets in {\color{blue} blue} state and $n$ (resp., $m$) is the number of vertices (resp.,
  edges) in the \textsf{VERTEX COVER} instance $H$. Thus, $\cs(M) = \mu  - \losses(M) - \ties(M)/2 \le \mu  -  2^{n-k-1}\cdot3^k\cdot10^m$. Moreover, $k \ge |C|$, where $C$ is a minimum vertex cover in $H$ (by Lemma~\ref{lem:vertex-cover}).

  Let us construct a matching $M_C$ where the vertex gadgets corresponding to the vertices in the minimum vertex cover $C$ are in
  {\color{blue} blue} state, those corresponding to the remaining vertices are in {\color{red} red} state, and for every edge
  $e = (i,j)$, if $\min\{i,j\}$ is in {\color{blue} blue} state then let $M_C \cap Y_e = F_e$, else $M_C \cap Y_e = L_e$. Then, the matching $M_C$ satisfies the conditions of Lemma~\ref{lem:Properties}. Therefore, by a similar argument as in the proof of Lemma~\ref{lem:vertex-cover}, we get that $\cs(M_C) =  \mu  - 2^{n-c-1}\cdot3^c\cdot10^m$, where $|C| = c$. Thus $\cs(M_C) > \cs(M)$ if $c < k$. Since $\cs(M)$ has to be the highest among all matchings, it follows that $c = k$. In other words, the 
  set of vertices whose gadgets are in {\color{blue} blue} state in $M$ constitute a minimum vertex cover in $H$.
\end{proof}

Finally, we show that our assumption that a Copeland winner does not use an inter-gadget edge always holds.
Thus Theorem~\ref{thm:Copeland_NPHard} stated in Section~\ref{sec:intro} follows.

\begin{restatable}{lemma}{lemmaStructureTwo}
\label{lem:structure2}
If $M$ is a Copeland winner in $G$, then $M$ does not use any inter-gadget edge.
\end{restatable}
\begin{proof}
The proof is similar to that of Lemma~\ref{lem:vertex-cover}. Below, we will outline the main steps involved in the proof.
Suppose, for contradiction, that $M$ uses an inter-gadget edge, say $(b_i,d_e)$ (see Fig.~\ref{fig:hardness2}). Let $\mathcal{X}$ denote the set of all vertex and edge gadgets in the matching instance. Consider a partitioning of $\mathcal{X}$ into \emph{single}, \emph{double}, \emph{triple}, and \emph{auxiliary} gadgets as follows:
\begin{itemize}
    \item Each \emph{double} (resp., \emph{triple}) gadget is a pair of an edge gadget and an adjacent vertex gadget $(Y_e,Z_i)$ (resp., a triple of an edge and two adjacent vertex gadgets $(Y_e,Z_i,Z_j)$) where $e = (i,j)$ is an edge such that $M$ contains an inter-gadget edge between $Y_e$ and $Z_i$ (resp., two inter-gadget edges between $Y_e$ and each of $Z_i$ and $Z_j$). While there is an edge gadget that shares an inter-gadget edge with one of (resp., both) its adjacent vertex gadgets, we make a double (resp., triple) gadget out of these two (resp., three) gadgets. Note that a vertex gadget can be included in at most one double/triple gadget in this manner.
    
    \item Next, if there is an edge gadget that is adjacent to two vertex gadgets both of which are in {\color{red} red} state and are still unclaimed by any edge gadget, then this edge gadget claims both these vertex gadgets; such a triple of edge gadget and its adjacent vertex gadgets is classified as an \emph{auxiliary} gadget. Note that such an edge gadget does not share an inter-gadget edge with either of the vertex gadgets.
    
    \item All remaining vertex and edge gadgets are classified as \emph{single} gadgets.
\end{itemize}
Observe that 
$$\losses(M) + \ties(M) \ge \Pi_{X\in \mathcal{X}}(\losses(M \cap X) + \ties(M \cap X)),$$ where $X$ denotes any single, double, triple or auxiliary gadget.

Consider any double or triple gadget $X$. Since the matching $M$ uses at least one inter-gadget edge $(b_i,d_e)$ in $X$, the vertex $a_i$ must be matched with either the vertex $b'_i$ or one of the vertices in $u_i^0,\dots,u_i^{99}$ (otherwise, if $a_i$ is unmatched, then $M$ will not be {\em Pareto optimal} 
and it is easy to see that every Copeland winner has to be Pareto optimal).\footnote{A matching $M$ is Pareto optimal if there is no matching $N$ such that at least one vertex is better off in $N$ than in $M$ and no vertex is worse off in $N$.} In both cases, the matching $\{(a_i,u_i^{k'}),(a'_i,b'_i)\}$ is tied with $M$ where 
$k' \in \{0,\ldots,99\}$. Thus, there are at least 100 matchings that defeat or tie with $M \cap (Y_e \cup Z_i)$. In other words, for every double or triple gadget $X$, the value of $\losses(M \cap X) + \ties(M \cap X)$ is at least 100. We therefore have
\begin{equation}
  \label{eqn:NoInterGadgetEdge_1}
  \losses(M) + \ties(M) \ \ge\ 2^{n'}\cdot 3^{n''}\cdot 10^{m'}\cdot 100^{t_2}\cdot100^{t_3}\cdot100^a,
\end{equation}  
where $n'$ (resp., $n''$) is the number of vertices present as single gadgets in {\color{red} red} (resp., {\color{blue} blue)} state, $m'$ is the number of edges that are present as single gadgets, $t_2$ is the number of
double gadgets, $t_3$ is the number of triple gadgets, and $a$ is the number of auxiliary gadgets in the aforementioned partition. As done in  Lemma~\ref{lem:vertex-cover}, we once again used Lemma~\ref{lem:ties} in bounding $\losses(M \cap Y_e) + \ties(M \cap Y_e)$ by 10 for 
every edge gadget $Y_e$ that is present as a single gadget, and used Observation~\ref{obs:Vertex_Gadget_Ties} in bounding $\losses(M \cap Z_i) + \ties(M \cap Z_i)$ by 2 (resp., 3) for 
every vertex gadget $Z_i$ (acting as single gadget) that is in {\color{red} red} (resp., {\color{blue} blue)} state. Additionally, we used Lemma~\ref{lem:blue} to obtain the corresponding bound for an auxiliary gadget.

We will now construct an alternative matching $M^*$ that has a higher Copeland score than $M$ to derive the desired contradiction. Starting with $M$, let us remove any inter-gadget edges from each double/triple gadget and convert both $M \cap Z_i$ and $M \cap Z_j$ in the triple gadget (or just $M \cap Z_i$ in case of a double gadget) to {\color{blue} blue} state and replace $M \cap Y_e$ with $F_e$. Additionally, for each auxiliary gadget $(Y_e,Z_i,Z_j)$, we convert both $M \cap Z_i$ and $M \cap Z_j$ to {\color{blue} blue} state and replace $M \cap Y_e$ with $F_e$. 

Note that for any edge $e = (i,j)$ in a single gadget, either $Z_i$ or $Z_j$ is now in {\color{blue} blue} state. If it is $\min\{i,j\}$ that is in {\color{blue} blue} state, then let $M^*\cap Y_{e} = F_{e}$, else let $M^*\cap Y_{e} = L_{e}$. The rest of the gadgets are in the same state as under $M$.

Notice that $M^*$ satisfies the conditions in Lemma~\ref{lem:Properties}. Thus, $\losses(M^*)=0$ and any matching that ties with $M^*$ must not use any inter-gadget edge. Therefore, the number of matchings that tie with $M^*$ across the entire graph $G$ is simply the product of matchings that tie with it on individual single, double, triple, and auxiliary gadgets. Hence,
\begin{align}
\label{eqn:NoInterGadgetEdge_2}  
  \losses(M^*) + \ties(M^*) \ & = \ties(M^*) \nonumber \\
  & = \ 2^{n'}\cdot 3^{n''} \cdot 10^{m'}\cdot 30^{t_2}\cdot90^{t_3}\cdot90^a,
\end{align}  
where $n',n'',m',t_2$, $t_3$ and $a$ are defined as in \eqref{eqn:NoInterGadgetEdge_1}. For any double or triple gadget, by Observation~\ref{obs:Vertex_Gadget_Ties} and Lemma~\ref{lem:10}, the entire contribution to $\ties(M^*)$ due to these two gadgets is 30 (or 90 in case of three gadgets) and to $\losses(M^*)$ is 0. Additionally, for an auxiliary gadget, the value of $\ties(M^*)$ is equal to $3\times3\times10=90$.

Comparing \eqref{eqn:NoInterGadgetEdge_1} and \eqref{eqn:NoInterGadgetEdge_2} along with the fact that $\losses(M) \ge 0 = \losses(M^*)$, we get that
$\cs(M^*) > \cs(M)$. 
Indeed, 
$$\losses(M^*) + \ties(M^*)/2 \ = \ 2^{n'-1}\cdot 3^{n''} \cdot 10^{m'}\cdot 30^{t_2}\cdot90^{t_3}\cdot90^a, \text{ and }$$
\begin{align*}
\losses(M) + \ties(M)/2 \ & \ge\ \left(\losses(M) + \ties(M)\right)/2 \\
& \ge\ 2^{n'-1}\cdot 3^{n''}\cdot 10^{m'}\cdot100^{t_2}\cdot100^{t_3}\cdot100^a.
\end{align*}
This contradicts the fact that $M$ is a Copeland winner. Thus, a Copeland winner must not use any inter-gadget edge.
\end{proof}

\begin{remark}
By reducing from a restricted version of \textsf{VERTEX COVER} on 3-regular graphs, which is also known to be $\mathsf{NP}$-hard~\cite{GJS76}, the intractability stated in Theorem~\ref{thm:Copeland_NPHard} can be shown to hold even when there are only a \emph{constant} number of neighbors per vertex.
\end{remark}

\section{Concluding Remarks}
We adopted a voting-theoretic perspective on the matching-under-preferences problem, and examined some existential and computational questions in the context of relaxing popularity. Though we know that a semi-popular matching always exists
and we showed an FPRAS to find an {\em almost} semi-popular matching, we do not know the computational complexity of finding 
an \emph{exact} semi-popular matching. The main open question here is settle this complexity. We also showed that it is $\mathsf{NP}$-hard to find a Copeland winner. Is there an FPRAS for an {\em approximate} Copeland winner?

Going forward, it will be very interesting to consider other voting rules that might facilitate tractability results while providing natural relaxations to well-studied solution concepts such as stability and popularity.

\section*{Acknowledgements}
We are grateful to the anonymous reviewers for their helpful comments. TK acknowledges support from project no. RTI4001 of the Department of Atomic Energy, Government of India. RV acknowledges support from SERB grant no. CRG/2022/002621 and DST INSPIRE grant no. DST/INSPIRE/04/2020/000107.

\clearpage

\clearpage
\begin{center}
    \text{\huge{Appendix}}
\end{center}

\section{Missing proofs from Section~\ref{sec:proof-Thm3}}
\label{sec:proofs}

\subsection*{Proof of Lemma~\ref{lem:10}}

\begin{proof}
We prove this lemma for the matching $F_e$. The proof for the matching $L_e$ is totally analogous.
Let us first fix the edges $(c_e,d_e), (c'_e,d'_e), (v_e,v'_e), (w_e,w'_e)$ and see what matchings that contain these 4 edges tie with $F_e$. 
There are 4 such matchings and the remaining edges in these matchings are the following four possibilities:

\begin{itemize}
    \item[1.] $(s_e,t''_e), (s'_e,t'_e), (s''_e,t_e)$
    \item[2.] $(s_e,t'_e),(s'_e,t_e),(s''_e,t''_e)$
    \item[3.] $(s'_e,t'_e),(s''_e,t''_e)$ 
    \item[4.] $(s_e,t'_e),(s''_e,t''_e)$
\end{itemize}

There are other matchings that contain the pair of edges $(c_e,d_e)$ and $(c'_e,d'_e)$ and are tied with $F_e$. There are again 4 such matchings and the
remaining edges in these matchings are the following four possibilities:

\begin{itemize}
    \item[5.] $(s_e,t'_e),(s''_e,t''_e),(v_e,v'_e),(t_e,w_e)$ 
    \item[6.] $(s'_e,t'_e),(s''_e,t''_e),(v_e,v'_e),(t_e,w_e)$ 
    \item[7.] $(s'_e,t'_e),(s''_e,t''_e), (s_e,v_e),(w_e,w'_e)$ 
    \item[8.] $(s'_e,t'_e),(s''_e,t''_e), (s_e,v_e), (t_e,w_e)$ 
\end{itemize}

Finally, the following 2 matchings are also tied with $F_e$. Note that both these matchings leave the vertex $d_e$ unmatched.
\begin{itemize}
    \item[9.] $\{(s'_e,t'_e),(s''_e,t''_e), (s_e,c_e), (v_e,v'_e), (w_e,w'_e), (c'_e,d'_e)\}$
    \item[10.] $\{(s'_e,t'_e),(s''_e,t''_e), (s_e,c_e), (v_e,v'_e), (t_e,w_e), (c'_e,d'_e)\}$
\end{itemize}

Thus there are 10 such matchings that tie with $F_e$ (note that $F_e$ is also included here). It can be verified that other than these 10 matchings, no other matching
in this subgraph is tied with $F_e$.

We now need to show that within the subgraph induced on $Y_e$, there is no matching that {\em defeats} $F_e$. The matching $F_e$ matches all vertices in
this subgraph and there is exactly one blocking edge wrt $F_e$ in this subgraph: this is the edge $(s''_e,t''_e)$. It follows from the characterization 
of popular matchings in
\cite{HK11} that a matching $N$ defeats $F_e$ if and only if $N\oplus F_e$ has an alternating cycle $C$ with the blocking edge $(s''_e,t''_e)$; moreover,
no edge in $C$ should be {\em negative}\footnote{An edge $(x,y)$ is negative with respect to matching $M$ if both $x$ and $y$ prefer their partners in $M$ to each other.}. 

It is easy to check there is no such alternating cycle wrt $F_e$. The absence of such a cycle implies that 
$F_e$ is popular in this subgraph. In other words, no matching in this subgraph defeats $F_e$.
\end{proof}

\subsection*{Proof of Lemma~\ref{lem:ties}}

\begin{proof}
Observe that it is enough to show this lemma for {\em Pareto optimal} matchings. 
We have already seen that $F_e$ (similarly, $L_e$) loses to {\em no} matching in the subgraph induced on $Y_e$. Hence for any matching 
$T_e \notin \{F_e,L_e\}$ in this subgraph, the 3 matchings $F_e, L_e$, and $T_e$ itself either defeat or tie with $T_e$. We need to show 
at least 7 more such matchings. We will divide this proof into three parts.

  \smallskip

  \noindent{\em (1)~Suppose neither $s_e$ nor $t_e$ is matched in $T_e$ to any of its top 3 neighbors.} So each of $s_e,t_e$ is either matched to
  one of its neighbors in the {\em triangle} that it is a part of or it is left {\em unmatched}.
  
  Observe that due to Pareto optimality of $T_e$, the edges $(s'_e,t'_e)$ and $(s''_e,t''_e)$ are in $T_e$.
  We will crucially use the triangles
  $\langle s_e,v_e,v'_e\rangle$ and  $\langle t_e,w_e,w'_e\rangle$ here. Let us use the phrase ``rotate the matching edge in a triangle'' to denote
  that we replace the edge of $T_e$ in this triangle with another edge that makes two of these vertices happy and one unhappy.

  \begin{itemize}
  \item We immediately get 3 matchings that defeat $T_e$ by rotating the matching edge in (i)~$\langle s_e,v_e,v'_e\rangle$,
    (ii)~$\langle t_e,w_e,w'_e\rangle$, and (iii)~both these triangles.
  \item Rotate the matching edge in $\langle s_e,v_e,v'_e\rangle$ and {\em free} the vertex $t_e$ from $\langle t_e,w_e,w'_e\rangle$, i.e.,
    if $t_e$ is matched in $T_e$ then replace $(t_e,\ast)$ with $(w_e,w'_e)$. We now get 3 matchings that tie with $T_e$ by each of these 3 choices:
    (i)~replacing the edge $(s'_e,t'_e)$ with $(s'_e,t_e)$, (ii)~replacing the edge $(s''_e,t''_e)$ with $(s''_e,t_e)$, and (iii)~replacing the edge
    $(c'_e,d'_e)$ with $(c'_e,t_e)$.

    Each of the operations~(i)-(iii) makes two vertices unhappy and one vertex $t_e$ happy. Since rotating the matching edge made 2 vertices 
    among $s_e,v_e,v'_e$ happy and one unhappy, everything summed together, it follows that each of these 3 matchings is tied with $T_e$.

  \item We do a mirror image of what we did in the earlier step, i.e., we rotate the matching edge in $\langle t_e,w_e,w'_e\rangle$ and
    {\em free} the vertex $s_e$ from $\langle s_e,v_e,v'_e\rangle$ and promote $s_e$ to one of its top 3 neighbors (at the cost of making 2 vertices
    unhappy).
  \end{itemize}

  \noindent{\em (2)~Suppose both $s_e$ and $t_e$ are matched in $T_e$ to one of their top 3 neighbors.}
  The Pareto optimal matching that results when $s_e$ (resp., $t_e$) is matched to its third choice neighbor is $F_e$ (resp., $L_e$).
  So we only have to consider the case when both $s_e$ and $t_e$ are matched to one of their top 2 neighbors. Observe that there has to
  be a blocking edge incident to the partner of $s_e$ (similarly, the partner of $t_e$).

  \begin{itemize}
  \item ``Unblock'' the blocking edge incident to $s_e$'s partner, i.e., if $(s_e,t'_e) \in M$ then replace this with $(s'_e,t'_e)$.
    This creates a more popular matching where $s_e$ is unmatched. To obtain another such matching, rotate the matching edge in
    $\langle s_e,v_e,v'_e\rangle$, i.e., replace $(v_e,v'_e)$ with $(s_e,v_e)$. Thus we get 2 matchings that are more popular than $T_e$.

  \item Run a ``mirror image'' of the above step with $t_e$ replacing $s_e$. This gives us 2 more matchings that are more popular than $T_e$.

  \item  Unblock the blocking edge incident to $s_e$'s partner {\em and} the blocking edge incident to $t_e$'s partner. This gives us one matching
    more popular than $M$. Now rotate the matching edge in $\langle s_e,v_e,v'_e\rangle$ -- this gives us another matching more popular than $M$.
    Instead, rotate the matching edge in $\langle t_e,w_e,w'_e\rangle$. Thus we can get 3 more matchings that are more popular than $M$.
  \end{itemize}

  \noindent{\em (3)~The remaining case.} So exactly one of $s_e,t_e$ is matched in $T_e$ to one of its top 3 neighbors.
  Without loss of generality, assume that it is $s_e$ that is matched to one of its top 3 neighbors.
  To begin with, let us assume that $(s_e,t'_e)$ and $(t_e,w_e)$ are in $T_e$.
  \begin{itemize}
  \item Rotate the matching edge in $\langle t_e,w_e,w'_e\rangle$, i.e., replace $(t_e,w_e)$ with $(t_e,w'_e)$. This gives us a matching
    more popular than $T_e$.
  \item Unblock the blocking edge incident to $s_e$'s partner, i.e., replace $(s_e,t'_e)$ with $(s'_e,t'_e)$. This gives us a second matching
    more popular than $T_e$. Also replace $(t_e,w_e)$ with $(t_e,w'_e)$. This gives us yet another matching more popular than $T_e$.
  \item Replace $(s_e,t'_e)$ with $(s'_e,t'_e)$ and rotate the matching edge in $\langle s_e,v_e,v'_e\rangle$, i.e., replace $(v_e,v'_e)$ with
    $(s_e,v_e)$. This gives us a 4th matching more popular than $T_e$. Now also rotate the matching edge in $\langle t_e,w_e,w'_e\rangle$. This gives us
    a 5th matching more popular than $T_e$.
  \item Replace the edge $(w_e,t_e)$ with $(s'_e,t_e)$. This gives us a 6th matching more popular than $T_e$ (this is not the same as
    $L_e$ since the edge $(w_e,w'_e) \in L_e$ while $w_e$ is unmatched here).
  \item Replace $(s_e,t'_e)$ with $(s'_e,t'_e)$ and replace $(t_e,w_e)$ with $(w_e,w'_e)$. This gives us a 7th matching that is tied with $T_e$
    since $s'_e,t'_e$, and $w'_e$ prefer this matching while $s_e,t_e,w_e$ prefer $T_e$.
  \end{itemize}
  
  Suppose $(s_e,c_e)$ and $(t_e,w_e)$ are in $T_e$. Then it is easy to see that corresponding to the first 5 matchings listed above, we again have 5 matchings 
  that defeat/tie with $M$. We can obtain two more matchings by unblocking the blocking edge incident to $c_e$, i.e., $(s_e,c_e)$ is replaced with $(c_e,d_e)$, 
  and (i)~replace $(s'_e,t'_e)$ and $(t_e,w_e)$ with $(s'_e,t_e)$ and $(w_e,w'_e)$ (ii)~replace $(s''_e,t''_e)$ and $(t_e,w_e)$ with $(s''_e,t_e)$ and $(w_e,w'_e)$.
  Observe that both these matchings are tied with $T_e$.
  
  When $(s_e,t''_e)$ and $(t_e,w_e)$ are in $T_e$, we can similarly show 7 matchings (other than $F_e,L_e,T_e)$ that defeat/tie with $T_e$. 
  When $(t_e,w'_e) \in T_e$ or when $t_e$ is unmatched in $T_e$, the above cases are analogous. This finishes the proof of Lemma~\ref{lem:ties}.
\end{proof}

\subsection*{Proof of Lemma~\ref{lem:structure1}}

The function $\vote$ will be useful to us here. For any vertex $x$ and any pair of matchings $S$ and $T$ in $G$, 
the vote of $x$ for $S$ versus $T$ can be defined as follows.
\begin{equation*} 
\mathrm{Let}\ \vote_x(S,T) = \begin{cases} 1   & \text{if $x$ prefers $S$ to $T$;}\\
	                     -1 &  \text{if $x$ prefers $T$ to $S$;}\\			
                              0 & \parbox[t]{0.55\linewidth}{otherwise,\ i.e., $x$ is indifferent between $S$ and $T$.}
\end{cases}
\end{equation*}

For any two matchings $S$ and $T$, observe that $$\Delta(S,T) = \sum_x \vote_x(S,T),$$ where the sum is over all vertices $x$ in $G$.

\begin{proof} (of Lemma~\ref{lem:structure1})
We know that the matching $M$ uses only intra-gadget edges. Let us start by showing that $(a_i,u^k_i) \notin M$ for any $k \in \{0,\ldots,99\}$.
Suppose not, i.e., let $(a_i,u^k_i) \in M$ for some $k \in \{0,\dots,99\}$. Then this means either $b_i$ or $b'_i$ is left unmatched in $M$.
\begin{enumerate}
    \item If $b_i$ is unmatched in $M$ then let $N = M \cup \{(a_i,b_i)\}\setminus \{(a_i,u^k_i)\}$.
    \item If $b'_i$ is unmatched in $M$ then let $N = M \cup \{(a_i,b'_i)\}\setminus \{(a_i,u^k_i)\}$.
\end{enumerate}

It is easy to check that for any matching $T$, we have $\Delta(N,T) \ge \Delta(M,T)$. In more detail, either
$\vote_{b_i}(N,T) - \vote_{b_i}(M,T) \ge 1$ or $\vote_{b'_i}(N,T) - \vote_{b'_i}(M,T) \ge 1$ since $N$ matches either $b_i$ or $b'_i$
to its top choice neighbor $a_i$ while $M$ leaves that vertex unmatched. We also know 
$\vote_{u^k_i}(N,T) - \vote_{u^k_i}(M,T) = -1$. For any other vertex $x$, we have $\vote_{x}(N,T) - \vote_{x}(M,T) \ge 0$. 

\smallskip

Thus we have $\Delta(N,T) \ge \Delta(M,T)$ for all matchings $T$ in $G$ and so $\cs(N) \ge \cs(M)$.
Moreover, $\Delta(N,M) > 0$ since two vertices prefer $N$ to $M$ (these are $a_i$ and one of $b_i,b'_i$)
while only $u^k_i$ prefers $M$ to $N$. Thus $\cs(N) > \cs(M)$, a contradiction to $M$ being a Copeland winner. 

Hence the matching $M$ has to use only the remaining 4 edges in $Z_i$. Note that a Copeland winner has to be Pareto optimal.
This implies there are only two choices for $M$ within the gadget $Z_i$: either $\{(a_i,b_i),(a'_i,b'_i)\} \subset M$ or $\{(a_i,b'_i),(a'_i,b_i)\} \subset M$.
\end{proof}

\subsection*{Proof of Lemma~\ref{lem:blue}}

\begin{proof}
  We will consider three cases here.

  \smallskip
  
  \noindent{\em (1) Either $(s_e,c_e)$ or $(t_e,c'_e)$ is in $M$.} In this case it is easy to see that there are at least 100
  matchings that tie with $M \cap (Y_e \cup Z_i \cup Z_j)$. Assume wlog that $(s_e,c_e) \in M$; so $d_e$ is left unmatched. 
  Consider the matchings obtained by replacing the edge $(a_i,b_i)$ with  the pair of edges $(d_e,b_i)$ and $(a_i,u^k_i)$. 
  Since $k \in \{0,\ldots,99\}$, there are 100 such matchings and each is tied with $M \cap (Y_e \cup Z_i \cup Z_j)$.

  \medskip

  \noindent{\em (2) Both $s_e$ and $t_e$ are matched to their top choice neighbors in $M$.} So $(s_e,t'_e)$ and $(s''_e,t_e)$ are in
  $M$. Consider the resulting matching obtained by replacing these two edges with $(s'_e,t'_e)$ and $(s''_e,t''_e)$ and replacing
  the edges $(c_e,d_e)$ and $(a_i,b_i)$ with $(s_e,c_e),(d_e,b_i)$, and $(a_i,u^k_i)$. It is easy to check that this matching is
  tied with $M$ since $s'_e, t'_e, s''_e, t''_e$, and $u^k_i$ prefer this matching to $M$ while $a_i,b_i,c_e,s_e$, and $t_e$
  prefer $M$ and all the other vertices are indifferent. Since $k \in \{0,\ldots,99\}$, there are at least 100 such matchings that are
  tied with $M$.

  \medskip
  
  \noindent{\em (3) The remaining case.} So either (i)~$s_e$ is matched to a neighbor worse than $c_e$ or (ii)~$t_e$ is matched to a neighbor
  worse than $c'_e$. Assume wlog that it is $s_e$ that is matched to a neighbor worse than $c_e$ (this includes the case 
  that $s_e$ is unmatched). We can assume $(t_e,c'_e) \notin M$ (see case~(1) above).

\begin{itemize}
\item Suppose $(s'_e,t_e)$ or $(s''_e,t_e)$ is in $M$. Consider the resulting matching obtained by replacing this edge with
  either $(s'_e,t'_e)$ or $(s''_e,t''_e)$ appropriately (and leaving $t_e$ unmatched) and replacing the edges $(c_e,d_e)$ and
  $(a_i,b_i)$ with $(s_e,c_e),(d_e,b_i)$, and $(a_i,u^k_i)$. In case $s_e$ is matched in $M$ to $v_e$ or $v'_e$, we replace this edge with 
  $(v_e,v'_e)$. For the sake of simplicity, we will assume that $(v_e,v'_e) \in M$. It is easy to check that this assumption is without loss of generality.

  The matching constructed above is tied with $M$ since $s_e, u^k_i$, and either $s'_e,t'_e$ or $s''_e,t''_e$ prefer this 
  matching to $M$ while $a_i,b_i,c_e$, and $t_e$ prefer $M$ and all the other vertices are indifferent. Since $k \in \{0,\ldots,99\}$, 
  there are at least 100 matchings that are tied with $M$.

\item If either $t_e$ was left unmatched in $M$ or matched to one
  of $w_e,w'_e$, then we can rotate the matching edge in the triangle $\langle t_e,w_e,w'_e\rangle$ to make two vertices happy and
  one unhappy (notice that the preferences within the triangle are cyclic). Again we assume wlog that $(v_e,v'_e) \in M$.

  Consider the resulting matching obtained by rotating the edge of $M$ in $\langle t_e,w_e,w'_e\rangle$ and
  replacing the edges $(c_e,d_e)$ and $(a_i,b_i)$ with $(s_e,c_e),(d_e,b_i)$, and $(a_i,u^k_i)$. 
  This matching is tied with $M$ since $s_e, u^k_i$, and two among $t_e,w_e,w'_e$ prefer this matching to $M$ 
  while $a_i,b_i,c_e$, and one of $t_e,w_e,w'_e$ prefer $M$ and all the other vertices are indifferent. 
  Since $k \in \{0,\ldots,99\}$, there are at least 100 matchings that are tied with $M$.
\end{itemize}
\end{proof}  

\subsection*{Proof of Lemma~\ref{lem:Properties}}

Before we prove this lemma, we describe the LP framework for popular matchings in roommates instances studied in \cite{Kav19}. 
It will be convenient to consider the graph $\tilde{G}$ which is $G$ augmented with self-loops, i.e., every
vertex considers itself its {\em bottom-most} acceptable choice, i.e., it finds being matched to itself more preferable than being unmatched.
We can henceforth restrict our attention to perfect matchings in $\tilde{G}$.

For any matching $N$ in $G$, define the matching $\tilde{N}$ in $\tilde{G}$ as $\tilde{N} \coloneqq N \cup \{(u,u): u$ is left unmatched in $N\}$. For any vertex $u$ and neighbors $x,y$ of $u$ in $G$, the function $\vote_u(x,y)$ will be useful to us. This is defined as follows:
\begin{equation*} 
\vote_u(x,y) = \begin{cases} 1   & \text{if $u$ prefers $x$ to $y$;}\\
	                     -1 &  \text{if $u$ prefers $y$ to $x$;}\\			
                              0 & \text{if $u$ is indifferent between $x$ and $y$.}
\end{cases}
\end{equation*}

To show property~(i) on the popularity of $M^*$, we will use the following weight function on the edge set of $\tilde{G}$.
For any edge $(u,v) \in E$, define $\wt(u,v) = \vote_u(v,\tilde{M}^*(u)) + \vote_v(u,\tilde{M}^*(v))$. 

So the weight of edge
$(u,v)$ is the sum of the votes of $u$ and $v$ for each other over their respective partners in $\tilde{M}^*$.
For any self-loop $(u,u)$, let $\wt(u,u) = \vote_u(u,\tilde{M}^*(u))$. So $\wt(u,u)$ is 0 if $\tilde{M}^*(u) = u$, else it is $-1$.
The definition of $\wt$ implies that for any perfect matching $\tilde{N}$ in $\tilde{G}$, we have $\wt(\tilde{N}) \coloneqq \sum_{(u,v) \in \tilde{N}} \wt(u,v) = \Delta(N,M^*)$. Thus, in order to show that $M^*$ is popular in $G$, it suffices to show that for every perfect matching $\tilde{N}$ in $\tilde{G}$, $\wt(\tilde{N}) \leq 0$. Below we will establish this inequality for a maximum weight perfect matching.

Consider the following LP that computes a max-weight perfect matching in $\tilde{G} = (V,\tilde{E})$ under the edge weight function $\wt$.
We will be using \ref{LP3} for analysis only -- we do not have to solve it.


\begin{linearprogram}
  {
    \label{LP3}
    \maximize{\sum_{e \in \tilde{E}} \wt(e)\cdot x_e}
  }
  \textstyle \sum\limits_{e \in \tilde{\delta}(u)}x_e \ & = \ \ 1 \ \ \forall\, u \in V\notag  \\
  \sum_{e \in E[S]}x_e\ & \le \ \ \lfloor|S|/2\rfloor\ \ \forall\, S \in \Omega \ \ \ \  \mathrm{and} \ \ \ \ x_e \ \ge \ \ 0\ \ \forall\, e \in \tilde{E}.\notag
\end{linearprogram}

Here $\tilde{E} = E \cup \{(u,u): u \in V\}$ and $\tilde{\delta}(u) = \delta(u)\cup\{(u,u)\}$, where $\delta(u) \subseteq E$ is the set of edges incident to vertex $u$. Also $\Omega$ is the collection of all odd-sized sets $S\subseteq V$. Note that $E[S]$ is the set of edges in $E$ 
with both endpoints in $S$.

\ref{LP3} computes $\max_{\tilde{N}}\wt(\tilde{N}) = \max_N\Delta(N,M^*)$ where
$\tilde{N}$ is any perfect matching in $\tilde{G}$. Thus $M^*$ is popular if and only if the optimal value of \ref{LP3} is 0. The dual of \ref{LP3} is given by \ref{LP4} below.
\begin{linearprogram}
  {
    \label{LP4}
    \minimize{\sum_{u \in V} y_u \ + \ \sum_{S\in\Omega}\lfloor\,|S|/2\,\rfloor\cdot z_S}
  }
  \textstyle y_u + y_v + \sum\limits_{S\in\Omega \, : \, u,v\in S} z_S \ & \ge \ \ \wt(u,v) \ \ \ \ \forall\, (u,v) \in E\notag  \\
  y_u \  & \ge \ \ \wt(u,u)\ \ \ \ \forall\, u \in V\ \ \ \ \mathrm{and} \ \ \ \ z_S \ \ge \ \ 0 \ \ \forall S \in \Omega\notag.
\end{linearprogram}

By LP-duality, $M^*$ is popular if and only there exists a feasible solution  $(\vec{y}, \vec{z})$ to \ref{LP4} such that $\sum_{u \in V} y_u \ + \ \sum_{S\in\Omega}\lfloor\,|S|/2\,\rfloor\cdot z_S = 0$. We will prove $M^*$'s popularity in $G$ by an appropriate assignment of values to $(\vec{y}, \vec{z})$. 

\smallskip

Regarding property~(ii), our dual solution will be an optimal solution to \ref{LP4}.
Moreover, every inter-gadget edge will be {\em slack} for this dual optimal solution. Hence complementary slackness conditions will imply that
$\wt(\tilde{N}) < 0$ for any matching $N$ in $G$ that contains an inter-gadget edge; in other words, $\Delta(N,M^*) < 0$ for such a matching $N$,
i.e., $N$ loses to $M^*$.

\begin{proof} (of Lemma~\ref{lem:Properties})
Let $z_S = 0$ for all $S \in \Omega$. We will assign $y$-values as described below.

Consider any edge $e = (i,j)$. Suppose the vertex gadget $Z_i$ is in {\color{blue} blue} state 
where $i = \min\{i,j\}$. So $M^* \cap Y_e = F_e$.

\begin{itemize}
    \item let $y_{s_e} = y_{t_e} = y_{s'_e} = -1$;
    \item let $y_{s''_e} = y_{t''_e} = y_{t'_e} = 1$;
    \item let $y_{v_e} = y_{w_e} = 1$ and $y_{v'_e} = y_{w'_e} = -1$;
    \item let $y_{c_e} = 1$ and $y_{d_e} = -1$ while $y_{c'_e} = -1$ and $y_{d'_e} = 1$.
\end{itemize}

Suppose the vertex gadget $Z_i$ is in {\color{red} red} state where $i = \min\{i,j\}$. So $M^* \cap Y_e = L_e$.
\begin{itemize}
    \item let $y_{s_e} = y_{t_e} = y_{t''_e} = -1$;
    \item let $y_{s''_e} = y_{s'_e} = y_{t'_e} = 1$;
    \item let $y_{v_e} = y_{w_e} = 1$ and $y_{v'_e} = y_{w'_e} = -1$;
    \item let $y_{c_e} = -1$ and $y_{d_e} = 1$ while $y_{c'_e} = 1$ and $y_{d'_e} = -1$.
\end{itemize}

For a vertex gadget $Z_i$ in {\color{blue} blue} state, we will set $y$-values as given below.
\begin{itemize}
    \item let $y_{a_i} = y_{b_i} = 1$; let $y_{a'_i} = y_{b'_i} = -1$;
    \item let $y_{u^k_i} = 0$ for all $k \in \{0,\ldots,99\}$.
\end{itemize}

For any vertex gadget $Z_i$ in {\color{red} red} state, we will set $y$-values as follows. 
\begin{itemize}
    \item let $y_{a_i} = y_{a'_i} = 1$; $y_{b_i} = y_{b'_i} = -1$;
    \item let $y_{u^k_i} = 0$ for all $k \in \{0,\ldots,99\}$.
\end{itemize}

It is easy to check that the above assignment of $y$-values along with $\vec{z} = \vec{0}$ is a feasible solution to \ref{LP4}. 
The self-loop constraints are easily seen to hold since we have $y_v \ge -1 = \wt(v,v)$ for all vertices $v$ other than the $u^k_i$ vertices.
Also, $y_{u^k_i} = 0 = \wt(u^k_i,u^k_i)$ for all $i$ and $k$.

We will check feasibility for the case when $Z_i$ is in {\color{blue} blue} state and $Z_j$ is in {\color{red} red} state. The other cases are totally analogous. For every edge $(u,v)$, we need to show that $y_u + y_v \ge \wt(u,v)$.
\begin{itemize}
\item $y_{a_i} + y_{b_i} = 2 = \wt(a_i,b_i)$ \ and \ $y_{a'_i} + y_{b'_i} = -2 = \wt(a'_i,b'_i)$.
\item $y_{a_i} + y_{b'_i} = 0 = \wt(a_i,b'_i)$ \ and \ $y_{a'_i} + y_{b_i} = 0 = \wt(a'_i,b_i)$.
\item $y_{a_i} + y_{u^k_i} = 1 > 0 = \wt(a_i,u^k_i)$ for $k \in \{0,\ldots,99\}$.
\item $y_{a_j} + y_{b_j} = 0 = \wt(a_j,b_j)$ \ and \ $y_{a'_j} + y_{b'_j} = 0 = \wt(a'_j,b'_j)$.
\item $y_{a_j} + y_{b'_j} = 0 = \wt(a_j,b'_j)$ \ and \ $y_{a'_j} + y_{b_j} = 0 = \wt(a'_j,b_j)$.
\item $y_{a_j} + y_{u^k_j} = 1 > 0 = \wt(a_j,u^k_j)$ for $k \in \{0,\ldots,99\}$.
\end{itemize}

When $Z_i$ is in {\color{blue} blue} state, recall that $M^* \cap Y_e = F_e$. So we have:
\begin{itemize}
\item $y_{s_e} + y_{t''_e} = 0 = \wt(s_e,t''_e)$ \ and \ $y_{s''_e} + y_{t_e} = 0 = \wt(s''_e,t_e)$.
\item $y_{s_e} + y_{t'_e} = 0 = \wt(s_e,t'_e)$ \ and \ $y_{s'_e} + y_{t_e} = -2 = \wt(s'_e,t_e)$.
\item $y_{s'_e} + y_{t'_e} = 0 = \wt(s'_e,t'_e)$ \ and \ $y_{s''_e} + y_{t''_e} = 2 = \wt(s''_e,t''_e)$.
\item $y_{s_e} + y_{v_e} = 0 = \wt(s_e,v_e)$ and $y_{s_e} + y_{v'_e} = -2 = \wt(s_e,v'_e)$.
\item $y_{t_e} + y_{w_e} = 0 = \wt(t_e,w_e)$ and $y_{t_e} + y_{w'_e} = -2 = \wt(t_e,w'_e)$.
\item $y_{v_e} + y_{v'_e} = 0 = \wt(v_e,v'_e)$ and $y_{w_e} + y_{w'_e} = 0 = \wt(w_e,w'_e)$.
\item $y_{s_e} + y_{c_e} = 0 = \wt(s_e,c_e)$ \ and \ $y_{t_e} + y_{c'_e} = -2 = \wt(t_e,c'_e)$.
\item $y_{c_e} + y_{d_e} = 0 = \wt(c_e,d_e)$ \ and \ $y_{c'_e} + y_{d'_e} = 0 = \wt(c'_e,d'_e)$.
\end{itemize}

Finally we will check that the inter-gadget edges are covered.
\begin{itemize}
    \item $y_{b_i} + y_{d_e} = 0 > - 1 = \wt(b_i,d_e)$ and $y_{b_j} + y_{d'_e} = 0 > - 1 = \wt(b_j,d'_e)$.
\end{itemize}

Observe that we have $y_u + y_v = 0$ for every edge $(u,v) \in M^*$ and $y_{u^k_i} = 0$ for every vertex $u^k_i$ left unmatched in $M^*$. Thus the objective function of \ref{LP4} with the above setting of $(\vec{y},\vec{z})$ evaluates to 0. Hence $M^*$ is popular in $G$. This finishes the proof of property~(i) of Lemma~\ref{lem:Properties}.

\medskip

We will now show property~(ii). Note that condition~2 in the statement of Lemma~\ref{lem:Properties} is necessary to show property~(ii): Indeed, we know from the proof of
Lemma~\ref{lem:blue} that without
this condition, i.e., if both $Z_i$ and $Z_j$ are in {\color{red} red} state for some edge $(i,j)$, there are matchings containing inter-gadget edges that do {\em not} lose to our matching.

We described above a vector $\vec{y}$ such that $(\vec{y},\vec{0})$ is a solution
to \ref{LP4} and we will now use complementary slackness to show property~(ii). 
Observe that $(\vec{y},\vec{0})$ is an optimal solution to \ref{LP4} since its objective function value is 0 which 
is the same as $\wt(M^*)$ since $\Delta(M^*,M^*) = 0$.

We will show that every inter-gadget edge is {\em slack} with respect to $(\vec{y},\vec{0})$. Notice that slackness of every inter-gadget edge with respect to $(\vec{y},\vec{0})$ implies, by complementary slackness, that any matching $N$ that includes an inter-gadget edge cannot be an optimal solution to \ref{LP3}. That is, $\wt(\tilde{N}) = \Delta(N,M^*) < 0$, or, in other words, $N$ loses to $M^*$. So for each edge $e = (i,j)$, where $i < j$, it suffices to check that the inter-gadget edges $(d_e,b_i)$ and $(d'_e,b_j)$ are slack. 

Whether $Z_i$ is in {\color{blue} blue} state or in {\color{red} red} state, we have $\wt(d_e,b_i) = -1$ since $b_i$ prefers $M^*(b_i) \in \{a_i,a'_i\}$
to $d_e$ while $d_e$ is indifferent between $M^*(d_e) = c_e$ and $b_i$.  Similarly, whether $Z_j$ is in {\color{blue} blue} state or in {\color{red} red} state, we have $\wt(d'_e,b_j) = -1$ since $b_j$ prefers $M^*(b_j) \in \{a_j,a'_j\}$ to $d'_e$ while $d'_e$ is indifferent between $M^*(d'_e) = c'_e$ and
$b_j$. 

Recall that $y_{b_i},y_{b_j},y_{d_e},y_{d'_e} \in \{\pm 1\}$.
So in the constraint for any inter-gadget edge in \ref{LP4}, the left side is an even number while the right side is $-1$. 
Thus every constraint in \ref{LP4} for an inter-gadget edge is slack. Hence any matching that contains an inter-gadget edges loses to $M^*$.
\end{proof}


\begin{thebibliography}{30}

\bibitem{APR05}
A.~Abdulkadiro{\u{g}}lu, P.~Pathak, and A.~E.~Roth.
\newblock The New York City High School Match.
\newblock {\em American Economic Review}, 95(2): 364--367, 2005.

\bibitem{APR+05}
A.~Abdulkadiro{\u{g}}lu, P.~Pathak, A.~E.~Roth, and T.~S{\"o}nmez.
\newblock The Boston Public School Match.
\newblock {\em American Economic Review}, 95(2): 368--371, 2005.

\bibitem{ABM06}
D.~J.~Abraham, P.~Biro, and D.~F.~Manlove.
\newblock Almost stable matchings in the Roommates problem.
\newblock {\em Proceedings of the 3rd Workshop on
Approximation and Online Algorithms (WAOA)}, 1--14, 2006.

\bibitem{BIM10}
P.~Biro, R.~W.~Irving, and D.~F.~Manlove.
\newblock Popular Matchings in the Marriage and Roommates Problems.
\newblock {\em Proceedings of the 7th International Conference on Algorithms and Complexity (CIAC)}, 97--108, 2010.

\bibitem{BB20}
F.~Brandt and M.~Bullinger.
  \newblock Finding and Recognizing Popular Coalition Structures.
  \newblock {\em Proceedings of the 19th International Conference on Autonomous Agents and MultiAgent Systems (AAMAS), 195--203, 2020}.
  
\bibitem{BCE+16}
F.~Brandt, V.~Conitzer, U.~Endriss, J.~Lang, and A.~D.~Procaccia.
\newblock Handbook of Computational Social Choice.
\newblock {\em Cambridge University Press}, 2016.

\bibitem{C00}
K.~Chung.
\newblock On the Existence of Stable Roommate Matchings.
\newblock {\em Games and Economic Behavior}, 33(2): 206--230, 2000.

\bibitem{C51}
A.~H.~Copeland.
\newblock A ``Reasonable'' Social Welfare Function.
\newblock Mimeo, University of Michigan, 1951.

\bibitem{wiki-copeland}
Copeland's Method.
\url{https://en.wikipedia.org/wiki/Copeland}

\bibitem{C17}
{\'A}.~Cseh.
\newblock Popular Matchings.
\newblock Chapter 6 in {\em Trends in Computational Social Choice}, 105(3), 2017.

\bibitem{CHK15} 
\'{A}.~Cseh, C.-C.~Huang, and T.~Kavitha. 
\newblock Popular Matchings with Two-Sided Preferences and One-Sided Ties.
\newblock {\em SIAM Journal on Discrete Mathematics}, 31(4): 2348--2377, 2017.

\bibitem{DM04}
P. Dasgupta and E. Maskin.
\newblock The Fairest Vote of All.
\newblock {\em Scientific American}, 290(3): 64--69, 2004.


\bibitem{FK22}
Y.~Faenza and T.~Kavitha.
\newblock Quasi-Popular Matchings, Optimality, and Extended Formulations.
\newblock {\em Mathematics of Operations Research}, 47(1): 427-457, 2022.

\bibitem{FKPZ19}
Y.~Faenza, T.~Kavitha, V.~Powers, and X.~Zhang.
\newblock Popular Matchings and Limits to Tractability.
\newblock {\em Proceedings of the 30th ACM-SIAM Symposium on Discrete Algorithms (SODA)}, 2790--2809, 2019.

\bibitem{FHS08}
P.~Faliszewski, E.~Hemaspaandra, H.~Schnoor.
\newblock Copeland Voting: Ties Matter.
\newblock {\em Proceedings of the 7th International Conference on Autonomous Agents and Multiagent Systems (AAMAS)}, 2: 983--990, 2008.

\bibitem{GS62}
D.~Gale and L.~S.~Shapley.
\newblock College Admissions and the Stability of Marriage.
\newblock {\em The American Mathematical Monthly}, 69(1): 9--15, 1962.

\bibitem{G75}
P.~G{\"a}rdenfors.
\newblock Match Making: Assignments Based on Bilateral Preferences.
\newblock {\em Behavioral Science}, 20(3): 166--173, 1975.

\bibitem{GJS76}
M.~R.~Garey, D.~S.~Johnson, L.~Stockmeyer.
\newblock Some Simplified NP-Complete Graph Problems.
\newblock {\em Theoretical Computer Science}, 1(3): 237--267, 1976.

\bibitem{GMSZ19}
S.~Gupta, P.~Misra, S.~Saurabh, and M.~Zehavi.
\newblock Popular Matching in Roommates Setting is NP-Hard.
\newblock {\em ACM Transactions on Computation Theory}, 13(2): Mar. 2021.

\bibitem{GI89}
D.~Gusfield and R.~W.~Irving.
\newblock The Stable Marriage Problem: Structure and Algorithms.
\newblock {\em MIT Press}, 1989.

\bibitem{HP01}
G.~H{\"a}gele and F.~Pukelsheim.
\newblock Llull's Writings on Electoral Systems.
\newblock {\em Studia Lulliana}, 41(97): 3--38, 2001.


\bibitem{H63}
W.~Hoeffding.
\newblock Probability Inequalities for Sums of Bounded Random Variables.
\newblock {\em The Collected Works of Wassily Hoeffding}, 409--426, 1994.

\bibitem{HK11}
C.-C.~Huang and T.~Kavitha.
\newblock Popular Matchings in the Stable Marriage Problem. 
\newblock {\em Information and Computation}, 222:~180--194, 2013.

\bibitem{Irv85}
R.~W.~Irving.
\newblock An Efficient Algorithm for the Stable Roommates Problem.
\newblock {\em Journal of Algorithms}, 6:~577--595, 1985. 
	
\bibitem{JS89}
  M.~Jerrum and A.~Sinclair.
  \newblock Approximating the Permanent.
  \newblock {\em SIAM Journal on Computing}, 18(6): 1149--1178, 1989.

\bibitem{Kav19}
T.~Kavitha.
\newblock Popular Roommates in Simply Exponential Time.
\newblock {\em Proceedings of the 39th IARCS Annual Conference on Foundations of Software Technology and Theoretical Computer Science (FSTTCS)}, 20:1--20:15, 2019.

\bibitem{K20}
T.~Kavitha.
\newblock Min-Cost Popular Matchings.
\newblock {\em Proceedings of the 40th Foundations of Software Technology and Theoretical Computer Science (FSTTCS)}, 25:1--25:17, 2020.



\bibitem{M13}
D.~Manlove.
\newblock Algorithmics of Matching under Preferences.
\newblock {\em World Scientific Publishing Company}, 2013.

\bibitem{M08}
R.~M.~McCutchen.
\newblock The Least-Unpopularity-Factor and Least-Unpopularity-Margin Criteria for Matching Problems with One-Sided Preferences.
\newblock {\em Proceedings of the Latin American Symposium on Theoretical Informatics (LATIN)}, 593--604, 2008.


\bibitem{MR95}
R.~Motwani and P.~Raghavan.
\newblock Randomized Algorithms.
\newblock {\em Cambridge University Press}, 1995.

\bibitem{Pacuit}
E.~Pacuit.
\newblock Voting Methods.
{\em The Stanford Encyclopedia of Philosophy.}
\newblock \url{https://plato.stanford.edu/archives/fall2019/entries/voting-methods}, 2019

\bibitem{PPR08}
N.~Perach, J.~Polak, and U.~Rothblum.
\newblock A Stable Matching Model with an Entrance Criterion Applied to the Assignment of Students to Dormitories at the Technion.
{\em International Journal of Game Theory}, 36(3), 519--535, 2008.

\bibitem{R84}
A.~E.~Roth.
\newblock The Evolution of the Labor Market for Medical Interns and Residents: A Case Study in Game Theory.
\newblock {\em Journal of Political Economy}, 92(6): 991--1016, 1984.

\bibitem{RP99}
\newblock A.~E.~Roth and E.~Peranson.
\newblock The Redesign of the Matching Market for American Physicians: Some Engineering Aspects of Economic Design.
\newblock {\em American Economic Review}, 89(4): 748--780, 1999.

\bibitem{RS92}
A.~E.~Roth and M.~Sotomayor.
\newblock Two-Sided Matching: A Study in Game-Theoretic Modeling and Analysis (Econometric Society Monographs).
\newblock {\em Cambridge University Press}, 1992.

\bibitem{VMAB16}
R.~Vaish, N.~Misra, S.~Agarwal, and A.~Blum.
\newblock On the Computational Hardness of Manipulating Pairwise Voting Rules.
\newblock {\em Proceedings of the 2016 International Conference on Autonomous Agents \& Multiagent Systems (AAMAS)}, 358--367, 2016.

\end{thebibliography}
\end{document}